\documentclass[letter, 10 pt, conference]{ieeeconf}  

\IEEEoverridecommandlockouts                              
\usepackage[utf8]{inputenc}
\usepackage[T1]{fontenc}
\usepackage{cite}
\usepackage{amsmath,amssymb,amsfonts}
\usepackage{algorithmic}
\usepackage{graphicx}
\usepackage{textcomp}
\usepackage{booktabs} 
\usepackage{float}
\usepackage{multirow}
\usepackage{subfigure}
\usepackage{cuted}
\usepackage{capt-of}
\usepackage{color, colortbl}
\definecolor{LightCyan}{rgb}{0.88,1,1}
\definecolor{LightGreen}{rgb}{0.7,1,0.7}
\definecolor{LightRed}{rgb}{1,0.4,0.5}
\definecolor{LightYellow}{rgb}{1, 0.850, 0.541}
\definecolor{Gray}{gray}{0.9}

\makeatletter
\let\NAT@parse\undefined
\makeatother
\usepackage{hyperref}  

\title{\LARGE \bf
Multimodal Interfaces for Effective Teleoperation
}

\author{Eleftherios Triantafyllidis$^{1}$, Christopher McGreavy$^{2}$, Jiacheng Gu$^{3}$ and Zhibin Li$^{4}$   
\thanks{All authors are with the School of Informatics, The University of Edinburgh, Scotland, United Kingdom.}
\thanks{$^{1}$E-mail: eleftherios.triantafyllidis@ed.ac.uk}%
\thanks{$^{2}$E-mail: c.mcgreavy@ed.ac.uk}%
\thanks{$^{3}$E-mail: j.gu@ed.ac.uk}%
\thanks{$^{4}$E-mail: zhibin.li@ed.ac.uk}%
\thanks{*This project was funded by the EPSRC Future AI and Robotics for Space (EP/R026092/1)}
}

\begin{document}

\maketitle
\thispagestyle{empty}
\pagestyle{empty}

\begin{strip}
    \centering
    \vspace{-10mm}
    \includegraphics[width = \textwidth,   height=5.8cm]{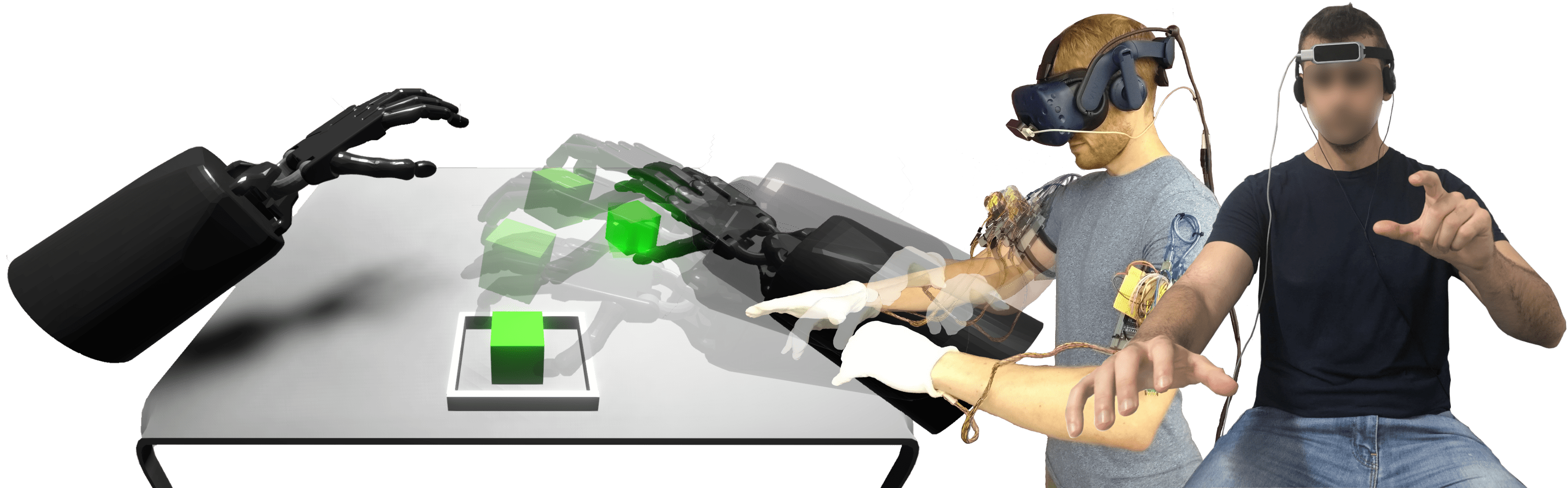}
    \vspace{-6mm}
     \captionof{figure}{Human operators with various interfaces and stimuli complete manipulation tasks of varying complexity in a first person view pick-and-place scenario. 
     }~\label{figure:teaser}
     \vspace{-6mm}
\end{strip}

\begin{abstract}
Research in multi-modal interfaces aims to provide solutions to immersion and increase overall human performance. 
A promising direction is combining auditory, visual and haptic interaction between the user and the simulated environment. 
However, no extensive comparisons exist to show how combining audiovisuohaptic interfaces affects human perception reflected on task performance. 
Our paper explores this idea. 
We present a thorough, full-factorial comparison of how all combinations of audio, visual and haptic interfaces affect performance during manipulation.
We evaluate how each interface combination affects performance in a study ($N=25$) consisting of manipulating tasks of varying difficulty.
Performance is assessed using both subjective, assessing cognitive workload and system usability, and objective measurements, incorporating time and spatial accuracy-based metrics. 
Results show that regardless of task complexity, using stereoscopic-vision with the VRHMD 
increased performance across all measurements by 40\% compared to monocular-vision from the display monitor. Using haptic feedback improved outcomes by 10\% and auditory feedback accounted for approximately 5\% improvement.
\end{abstract}

\section{INTRODUCTION}
\PARstart{T}{he} growth of virtual reality, robotics and networking technologies have spiked in recent years.
This has led to an increase in teleoperation research -- allowing humans the ability to remotely inhabit a foreign body, e.g. a robot as an avatar to complete a task \cite{10.1007/3-540-44589-7_1}. With the recent outbreak of pandemics, remote robotic control and telepresence systems, have become more important than ever.

Teleoperation delegates the high-level control of a robot to a remote human operator, thus combining the human instinct and the computational as well as the physical capabilities of robots. Humans are highly adaptable experts in motion control, constituting teleoperation a useful tool to help robots complete tasks in novel and dynamic environments.
During a teleoperation task, the robot's performance is dictated by the controls being sent by the human.
So how can we maximize human perception and thus maximise performance during task supervision?

The actions between an operator and a remote robotic system are physically detached from another constituting the overall experience unnatural.
This implies that policies which humans use to control their own bodies, may not directly translate into effective control of a foreign body, which can lead to poor performance.
To mitigate this, we can maximise feelings of immersion and by extent task performance in humans so that the foreign body feels more like their own.
This can lead to improved performance when controlling another body in a remote environment \cite{Sigrist2013, 4343985}. 
Increasing immersion, translates in increased performance \cite{yanco_drury_2004, 1308838, Jennett:2008:MDE:1393652.1393920}.

To increase the feeling of immersion and thus performance, we can alter the way in which the human interacts with their avatar. 
In a primary setting, users can interact with their surrounding virtual setting, by using a monocular monitor to give them a visual representation of the environment in which they are operating, which may not necessarily lead to high levels of immersion by itself.
Using a virtual reality device could lead to increased performance because it offers richer visual information, particularly attributed to stereoscopic depth \cite{martins_ventura_2009, doi:10.1177/154193129503902006}.
Stimulating other senses however, for example, using auditory and haptic feedback to superimpose information may also affect performance. 

Previous work has compared the effect of combining some sensory interfaces.
However, exhaustive comparisons, to the best of our knowledge, have yet not been made between visual, haptic and auditory sensory modalities and how combining these affects task performance of varying complexity.
Our work aims to address this. 

We use a pick and place task to compare the effects of these sensory interfaces on task performance. 
The setup for this task can be seen in \autoref{figure:teaser}.
The pick and place task is set in a virtual environment with different objects types, sizes and pick and place distances.
We compare all combinations of visual (monocular or VR), auditory (presence or absence) and haptic (presence or absence) feedback.
Changing these factors affects the difficulty of the task and we present a detailed analysis on how each combination of sensory inputs affects task performance.

Our study provides evidence to support a recommendation for the best performing combination of sensory interface in manipulation tasks with varying complexity. 
By incorporating both subjective and objective measurements, we determine which combination offers the best performance for a given task.
Throughout this paper, we present how we conduct, evaluate and analyse our experiments.

Contributions of our work include:
\begin{itemize}
\setlength
    \item A unique and reproducible interface which allows various combinations of sensory feedback for performing various tasks under different settings,
    \item A low cost hardware and simple software approach in designing an effective vibrotactile haptic data glove,
    \item A virtual reality environment with high-fidelity physics simulation (friction, collision, contact forces) to closely resemble real-world interaction and make the best use of existing human motor skills,
    \item A concrete experimental design that can be used to test the effectiveness of new emerging technologies,
    \item To the best of our knowledge, this is the first exhaustive comparison of its kind between all combinations of visual, auditory and haptic interfaces during manipulation tasks of increasing difficulty.
\end{itemize}

\section{Related Work}
In this section, we discuss the previous work regarding the effectiveness of multi-sensory interfaces on immersion and performance, object interaction and manipulation.
We group these studies in the individual sensory modalities for clarity and identify gaps in current knowledge. 

\subsection{Multi-Modal Interfaces}
When operators embody a remote robot or are subjected to a virtual environment for training purposes, using only a visual monocular monitor, they can only experience that remote environment visually.
By adding multiple modalities, it was found that the workload of the visual cortex can be reduced, awareness may be increased and thus task performance can be improved \cite{Lathan:2002:EOS:638095.638100, doi:10.1080/10447319609526138}.
But when using multi-modal interfaces, synchronisation is important. 
If signals of different modalities are out-of-sync, overall spatial and temporal immersion is reduced, effectively nullifying the benefits of using multi-modal interfaces \cite{popescu2002multimodal, richard1994comparison}.

Furthermore, sensory feedback strategies need to be made prior to the implementation of a specific sensory channel. In most cases, the design decisions of one type of sensory feedback may be achieved via either a continuous manner, i.e. concurrent feedback, or after a desired event, i.e. terminal feedback \cite{Sigrist2013}. 

This study focuses on audiovisuohaptic interfaces,
due to vision, hearing and touch being the highest developed and contributing the most in embodiment
\cite{popescu2002multimodal, Heilig1992ELCD, 4343985}, among all human senses. We present the previous work on visual, audio and haptic interfaces in the following sections. \\

\subsubsection{Visual Cues}
Most research in this area has focused on the effect of visual interfaces between the human and the avatar.
The dominance of vision in the sensory system is well supported \cite{Rock594,10.2307/40063340, MCINTIRE201418}, contributing to around 70\% of overall human perception \cite{Heilig1992ELCD}.
Providing thus visual information in the best form is of vital importance.

The two primary sources of a visual interface are standard monocular monitors and VRHMD, which provides stereo vision.
During a target detection task in Unmanned Aerial Vehicles (UAVs), there were no significant differences in performance between the two \cite{doi:10.1177/1541931213601863}, with the VRHMD even causing motion sickness potentially attributed to the illusion of self-motion of the vehicle.
This is known as vection \cite{definition_vection} and is a common complaint among VRHMD users in non-static situations.
Since this is still an open problem, our study limits self-motion and compares the effectiveness of both displays in static scenarios.

Though VRHMDs have drawbacks, they also offer many benefits over monocular screens.
They offer better depth perception and environmental awareness than standard monitors \cite{380802}. This is of importance as studies have shown that humans overestimate their ability to perceive depth in virtual environments \cite{doi:10.1177/154193129503902006,Witmer:1998:JPT:1246749.1246755, 7164348}. As such, increased depth information leads to reduced collisions with the surrounding environment and better performance during highly dexterous manipulation tasks \cite{martins_ventura_2009,scribner1998effect}. It is important, however, how the superimposition of information is delivered to the operator. One study showed that constantly providing feedback can be counterproductive both in user preference and time efficiency compared to providing feedback at the end of a task \cite{Shang:2019:ECS:3308557.3308675}.

Providing a larger field of view can also result in increased performance and environmental awareness \cite{scribner1998effect, doi:10.1177/154193120004403624,Johnson:2015:YSM:2702123.2702526}, but can decrease usability and increase perceived difficulty i.e. workload \cite{Johnson:2015:YSM:2702123.2702526}. 

\vspace{20mm}

\subsubsection{Auditory Cues}
Supplementing vision with auditory information can lead to increased operator awareness, especially during high visual load \cite{shilling2002virtual, Sigrist2013} and can reduce mental workload, correlated to fewer accidents and better performance \cite{nagai_kimura_tsuchiya_iida_2000}.
Audiovisual interfaces also contributed to intuitive control of a humanoid during manipulation tasks \cite{doi:10.1163/156855303764018468}.

Extra information, such as alarms and alerts can be superimposed on a visual display, but by presenting them via an audio interface can decrease distraction \cite{Secoli2010EffectOV}.
Operators can also use auditory information to localise the sources of sounds, which is useful when FOV is limited \cite{simpson2013spatial}. 

Further studies also show that controlling auditory pitch may influence object clearance, limited during human walking, with results indicating that participants indeed benefited from such sound sonification \cite{erni2001obstacle, wellner2008evaluation}. This suggests that auditory information may provide a richer environmental experience and be a valuable supplement to just relying on vision. \\

\subsubsection{Somatosensory Cues}
Tactile feedback can also augment visual information.
Communicating spatial alerts via somatosensory means can signal warnings without overloading visual pathways \cite{Cholewiak2000, Rochlis2000ATD}.
Manipulation, in particular, can be improved by adding tactile feedback \cite{962082, 10.1007/3-540-44589-7_1, 8446403} and can result in better performance \cite{8446227}. 

For diagnostic surgery simulators using virtual reality, complex and sophisticated tactile approaches for force feedback have been developed to allow realistic reaction forces for deformable objects such as soft tissue \cite{844824}. Further research in kinesthetic force feedback, has shown some advantages over lower cost approaches \cite{10.1145/3332165.3347891, 10.1145/2858036.2858487}, particularly due to being able to constrain the grasp motion of users hands, based on the virtual object they are holding \cite{choi_hawkes_christensen_ploch_follmer_2016}. 

However, providing high resolution haptic feedback alone does not necessarily guarantee an increase in task performance \cite{BergerUncannyValleyHaptics}. Using only vibration feedback can increase spatial awareness for non-deformable i.e. rigid objects \cite{vibration_proximity}.
Outputting vibrations which are proportional to the force applied by the robot, also leads to improved performance \cite{Murray:2003:PCT:782655.782658}. 
We use a similar approach. \\

\subsubsection{Audiovisuohaptic Multi-Modal Interfaces}
A combination of all three modalities may also be effective in improving performance. One study hypothesises that audiovisuohaptic interfaces may increase task performance as the task gets gradually more difficult \cite{Sigrist2013}, but this is untested. 

On one hand, an audiovisuohaptic interface did not significantly increase performance during a teleoperated navigation task \cite{Lathan:2002:EOS:638095.638100}, but operator spatial ability and subjective performance did increase compared to using fewer interfaces. In another study, an audiovisuohaptic interface was implemented to test for performance in visual throwing tasks \cite{Frid2018}. While not exhausting all comparisons of the interface or implementing varying task complexities, their results show that point-based haptic devices and moreover auditory feedback did not contribute to significantly improved task performance.

A meta-analysis of 45 studies showed that by supplementing visual information with either auditory or somatosensory (via vibrotactile cues), increased overall performance \cite{Burke:2006:CEV:1180995.1181017}.
However, no extensive comparison has been conducted on how combining all three modalities affects immersion and by extent task performance reflected on higher levels of complexity. Our study aims to address this.

\subsection{Object Interaction and Manipulation}
To compare the effect of visual, auditory and haptic feedback on task performance, we must first define a task. 
We chose to measure the effect of these interfaces on manipulation tasks of different difficulties. 
Manipulation is a suitable choice, since it involves coarse and fine motor movements, depending on the object being grasped. 

The Southampton Hand Assessment Procedure (SHAP) \cite{light_chappell_kyberd_2002}, defines six clinically validated grasping classifications to test hand function. 
This comprises the entire range of human hand motion from fine to coarse manipulation.
One study even addressed all the possible different grasping techniques a human can initiate with an object by implementing the SHAP in the physics engine MuJoCo, however, no comparison between the sensory modalities was drawn \cite{MuJoCoHaptix}. We are undoubtedly inspired by the aforementioned study. During our experiments, we use a range of different objects and sizes.
By doing this we can examine the effect of combining sensory interfaces on the performance of different levels of human motor skill during object manipulation and interaction.

Our aim is to increase task performance by improving immersion. 
However, immersion is a complex phenomenon which can be negatively influenced by the so-called "Uncanny Valley" -- a break in immersion when an artificial being appears \textit{too} realistic, causing negative responses towards it \cite{DefinitionUncannyValley}. 

More relevant to this study is the "Uncanny Valley of Haptics", which has a similar effect when haptic feedback does not coincide with other sensory feedback and reduces the perception of realism \cite{BergerUncannyValleyHaptics}.
Neuroimaging studies support this concept, showing that visual and haptic activation overlaps in the occipital lobe \cite{IsNeoCortexEssentiallyMultisensory, Amedi2001VisuohapticOA, JamesThomasKeithHapticActivatesVisualAreas, SathianZangaldzeVisualCortexToTactilePerception}.
We aim to investigate if the simultaneous presence of both modalities increases performance.

\section{Hypotheses}
The following hypotheses are formed from our review, while primarily hypothesizing that an audiovisuohaptic multi-modal interface will prove to be significantly more effective when subjected to higher task complexity, compared to fewer modalities present or the minimal representation of these. 

\vspace{30mm}

\textbf{Hypothesis 1:} There will be lower perceived cognitive workload corresponding to higher performance with (a) the stereoscopic VRHMD than with the monocular display monitor, (b) presence of somatosensory feedback than absence and finally (c) presence of auditory feedback than the absence of it. \\

\textbf{Hypothesis 2:} There will be higher perceived system usability corresponding to higher performance with (a) the stereoscopic VRHMD than with the monocular display monitor, (b) presence of somatosensory feedback than absence and finally (c) presence of auditory feedback than the absence of it. \\

\textbf{Hypothesis 3:} Faster performance corresponding to less placement and completion time will be observed with (a) the stereoscopic VRHMD than with the monocular display monitor, (b) presence of somatosensory feedback than absence and finally (c) presence of auditory feedback than the absence of it. \\

\textbf{Hypothesis 4:} Better depth estimation with less distance error to target, will be measured in the order of interface conditions incorporating (a) the stereoscopic VRHMD than with the monocular display monitor, (b) presence of somatosensory feedback than absence and finally (c) presence of auditory feedback than absence. \\

\textbf{Hypothesis 5:} Higher placement precision, including higher spatial position and orientation accuracy, will be measured in the order of interface conditions incorporating (a) the stereoscopic VRHMD than with the monocular display monitor, (b) presence of somatosensory feedback than absence and finally (c) presence of auditory feedback than the absence of it. \\

\section{Methodology}
This section describes the key hardware and software components in our study. First, to test our hypotheses, we designed a series of experiments. 
During these experiments, participants performed a pick and place task under various conditions.
All possible combinations of a visual, auditory and haptic interface are assessed.
Each modality has two modes, as detailed in \autoref{table:SensoryModalities}, providing a full factorial study.
\begin{table} [H]
\centering
\begin{scriptsize}
\begin{tabular}{lcccccc}
    \toprule
     &
      \multicolumn{2}{c}{\textbf{Vision}} &
      \multicolumn{2}{c}{\textbf{Audition}} &
      \multicolumn{2}{c}{\textbf{Haptics}} \\
      \cmidrule(lr){2-3}
      \cmidrule(lr){4-5}
      \cmidrule(lr){6-7}
      & {Monitor} & {VRHMD} & {Absence} & {Presence} & {Absence} & {Presence} \\
      \midrule
       \rowcolor{Gray} C1  & X &  & X &  & X &  \\
       C2  & X &  & X &  &  & X  \\
       \rowcolor{Gray} C3   & X &  &  & X  & X &  \\ 
       C4   & X &  &  & X  &  &  X \\ 
       \rowcolor{Gray} C5    &   & X & X  &  & X  & \\
       C6    &   & X & X  &  &   & X \\
       \rowcolor{Gray} C7   &   & X &  & X  & X &  \\
       C8    &   & X &  & X  & & X \\
    \bottomrule
  \end{tabular}
    \end{scriptsize}
  \caption{The multi-modal interface broken down into the $2^3$ possible combinations of visual, auditory and haptic feedback.}
  \label{table:SensoryModalities}
  \vspace{-2.5mm}
\end{table}
Audition and haptics can be on/off, whereas vision can be either represented by a monocular i.e. display monitor or a stereoscopic i.e. VRHMD.
All combinations of these modalities amount to $2^3$ combinations. 
Assessment of performance is achieved via both objective and subjective metrics. Participants completed manipulation tasks under each of the above conditions. 

\subsection{Participants}
A total of ($N=25$) participants were recruited in this study via an advertisement at the University of Edinburgh. 
Ages ranged from 21 to 44 ($M=26.36$, $SD=4.829$, $Mdn=26$), with 6 females and 19 males.
Each had healthy hand control and normal/corrected vision.
A 30-minute interactive experience using the VRMHD was given as compensation.

\subsection{Equipment and System Setup}
For visual feedback, a computer monitor was used for the monocular condition and a VRHMD for stereoscopic vision.
The monitor was a 27 inch HP Elite IPS display, with 2560 x 1440 resolution and 60Hz refresh rate placed 75-100cm from the participant.
The VRHMD was HTC Vive Pro with 3.5 inch AMOLED screen at 2880 x 1600 resolution (1440 x 1600 pixels per eye), 90Hz refresh rate and $110^{\circ}$ FOV. 
High-resolution displays were chosen to limit distance overestimation and a degraded longitudinal control \cite{10.1080/0014013032000121624}. 
An NVIDIA 2080 RTX Ti was used to ensure consistent frame-rates. 

Two stereo headsets provided the audio interface.
One was integrated onto the HTC Vive VRHMD for stereo conditions.
The other was separately attached during the display monitor monocular conditions. 
Audio quality was at 16 bit, 44100 Hz.

To provide haptic feedback, we constructed a custom haptic glove inspired by \cite{HapticFeedbackVibration}, which incorporated a vibration motor on the thumb and index finger of the glove. The vibration intensity in their study was accomplished and influenced by the proportion of the size of the virtual object the user was colliding and touching with. While their approach indeed shows a promising step towards immersive experiences in the branch of entertainment, we took their method a step forward by incorporating physical properties including kinetic energy and object penetration for manipulation scenarios detailed further along this paper, specifically in the methodology section. In the construction of our custom glove, 15 coin-vibration motors were used, with DC 3V 70mA 12000 RPM. 
Two motors were placed on each finger (proximal \& distal phalanges). 
Five motors were placed on the palm.
Wireless communication between the virtual environment and the glove ensured free movement. 
This was achieved using a Bluetooth transceiver for each glove.

We chose to use vibrotactile stimulation rather than force feedback for its lower cost and certain advantages over force feedback.
Preliminary findings indicate that force feedback is only more beneficial than vibrotactile stimulation when presented at a high-resolution \cite{BergerUncannyValleyHaptics,10.1007/978-3-642-39405-8_28}.
But this increases cost and size. Air jet driven approaches exist for force feedback, while these show significant effectiveness, they nonetheless require large space and pose a substantially higher cost compared to vibration approaches \cite{1381224}.
Vibratory feedback, however, can be more beneficial than force feedback in direct manipulation tasks such as ours \cite{Kontarinis:1995:TDV:2870994.2870999}.
Overall vibrotactile stimulation is shown to be an effective substitute for force feedback according to another study \cite{doi:10.1162/pres.1993.2.4.344}.

The manipulation task for this study was performed by mapping the user's hand movements to an anthropologically human-robot hand in the simulation environment.
To capture hand movements, we used the Leap Motion Hand Controller (LMHC). 
This uses a stereo camera system and infrared LEDS to capture hand motions.
In all conditions, the device was fixed to the participant's forehead, either by a strap or on the front of the VRHMD. 
The LMHC was able to track the haptic gloves, as anthropomorphic features were retained. 

\subsection{Software and Simulation Environment Setup}
\begin{figure} [H]
\centering
  \includegraphics[width=\columnwidth]{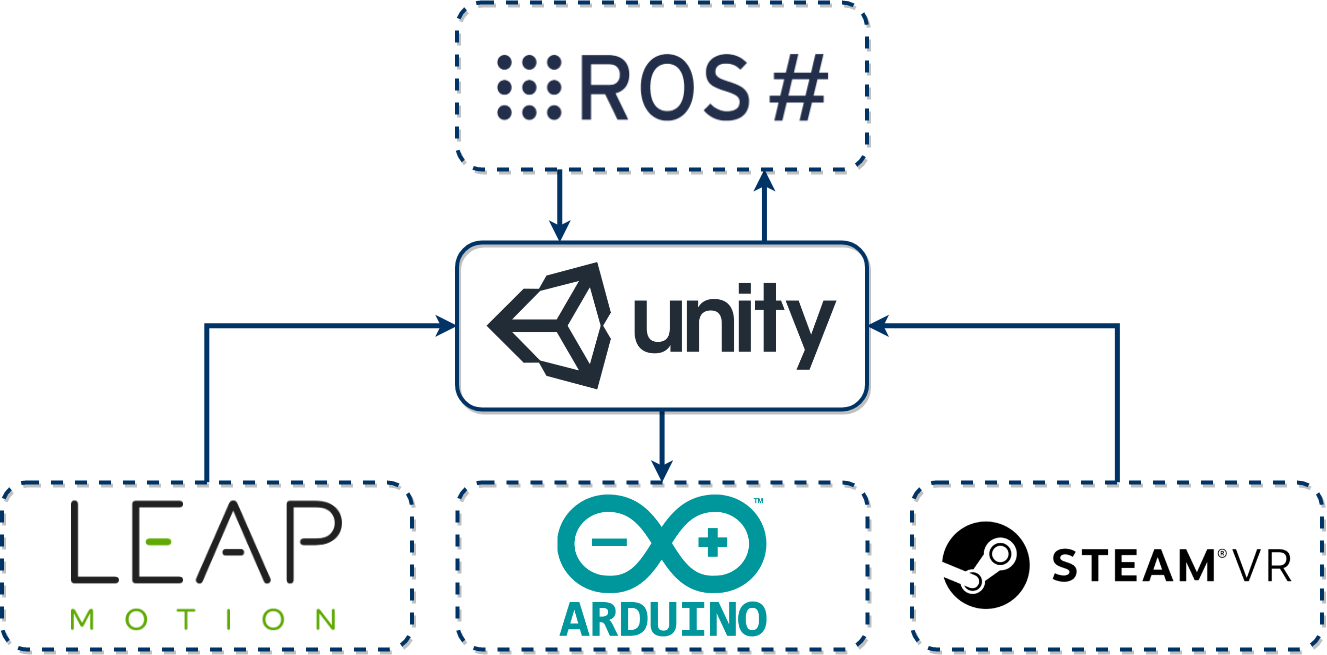}
      \vspace{-2.5mm}
  \caption{Diagram of the simulation setup with all the software plugins used.}~\label{fig:SoftwarePlugins}
\end{figure}

In our experiments, the participant's conducted manipulation tasks in a virtual environment.
As such, this study required a virtual environment which was connected to the hardware.
The relationship of these components is shown in \autoref{fig:SoftwarePlugins}.
The Unity3D engine was used as the core of our virtual environment.
Two Shadow robotic hands acted as teleoperated manipulators.
Physics simulations of the environment used the Unity3D engine, whereas robotic hand physics were handled by the ROS-Sharp physics engine.
Unity obtained hand positions from the LMHC via the Leap Motion SDK.
A plugin was developed to communicate between the Unity environment and haptic gloves via a Bluetooth module on the glove's Arduino controllers. 

\subsection{Hand Manipulation and Control}
\begin{figure} [H]
\centering
  \includegraphics[width=\columnwidth]{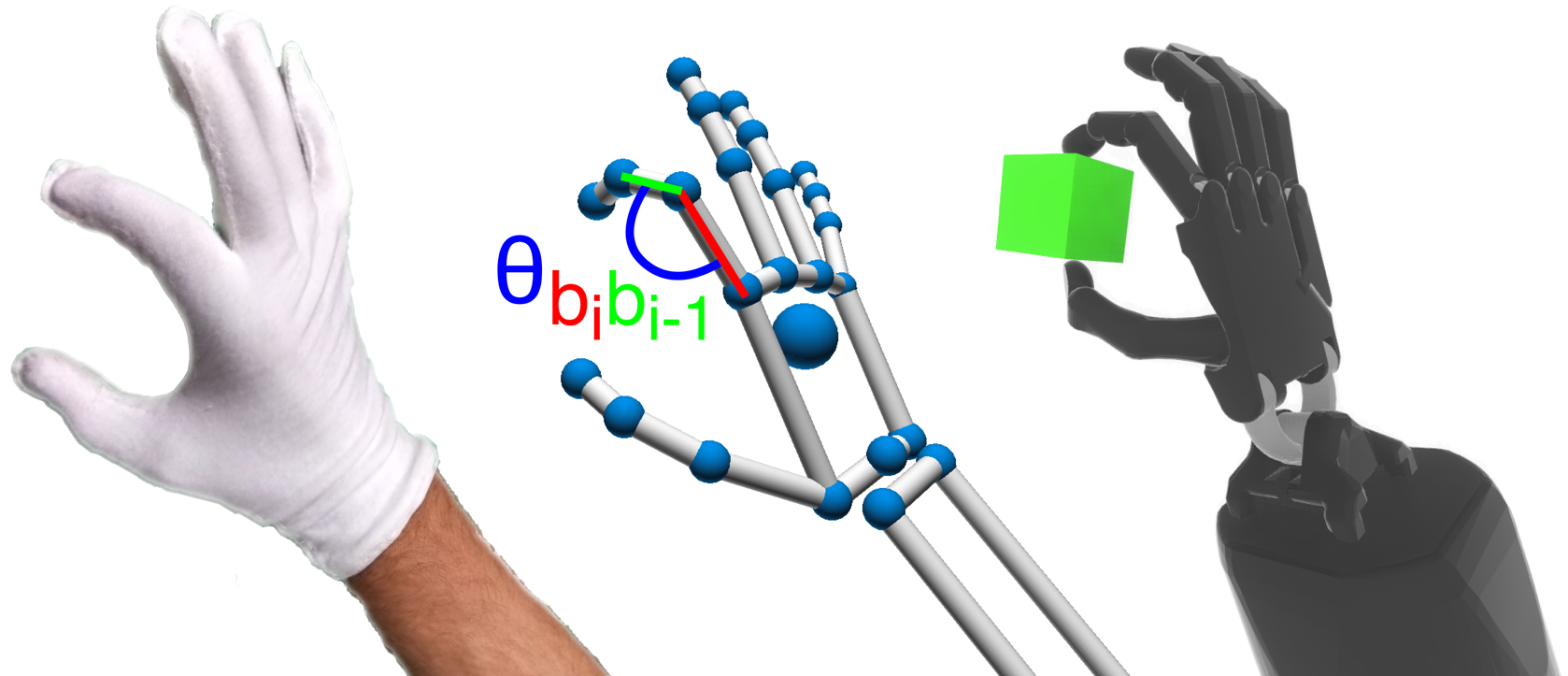}
        \vspace{-7.5mm}
  \caption{Hand control approach through direct joint angle re-targeting from our custom haptic glove to the final robotic hand.
}~\label{fig:HandControlAngle}
\end{figure}

The Leap Motion SDK outputs Cartesian joint positions in world frame, but joint angles are required to control the virtual hand.
This translation was made by calculating the angle $\theta_{}$ between a joint $\vec{b_{i-1}}$ and its parent $\vec{b_{i}}$.
\begin{equation}
\theta = \arccos \left ( \frac{\vec{b_{i}} \cdot \vec{b_{i-1}}}{\left \| \vec{b_{i}} \right \| \left \| \vec{b_{i-1}} \right \|} \right ) ,
\end{equation}
A Proportional Derivative (PD) controller was used to control the joints. 
Each joint has one PD controller, formulated as follows for each timestep $t$:
\begin{equation}
u(t) = K_P \cdot e(t) + K_D \cdot \dot{e}(t),
\end{equation}
where $u(t)$ is the angular velocity control signal sent to the Shadow hand joints.
$e(t) = q_{h}(t) - q_{r}(t)$ is the current position error between the human joint and the robot joint and $\dot{e(t)} = \dot{q}_{ref} - \dot{q}_{r}(t)$ is the velocity error between the robot and the desired velocity, which here is set to zero.
$K_P$ and $K_D$ are the gains which were tuned such that human and robot motion matched as accurately as possible. Depicted in \autoref{fig:HandControlAngle}.

\subsection{Sensory Interface Design}
\subsubsection{Visual Stimulation}
We compare monocular and stereo feedback in our experiments using a generic display monitor and a VRMHD respectively. In addition however to providing visual disparity, the VRHMD also allows users to control the viewpoint in the virtual environment by moving their head.
To conduct a fair experiment, we allow participants to change their viewpoint when using the monitor by using a computer keyboard using standard gaming keybindings, retain the optical hand controller consistently in a head-mounted state as well as using a monitor of similar resolution to the VRHMD. Acclimatization to these controls and technologies were allowed prior to commencing the experiments, detailed further along this work. \\

\subsubsection{Auditory Stimulation}
We hypothesize that auditory feedback will contribute to increased performance. Every day sound effects, "that make sense", were used to investigate how sound may compensate the superimposition of visual information without requiring prior context or explaining these to the participants i.e. would be inherently perceived as a substitute to text. Audio feedback is given in two situations.

Firstly, warnings and notifications were given via audio.
A high-pitched alarm sound warned of imminent collisions between the robotic hands and the environment.
A siren alarm sound on the other hand indicated time was running low.
A successful "ding" indicated that at least part of an object had been placed inside the target volume irrespective of the placement accuracy.

Auditory feedback also relayed the sounds of interactions in the environment.
Picking up, dropping or placing an object produced realistic bump and scrape sounds one would expect when interacting with real objects. \\

\subsubsection{Somatosensory Stimulation}
Vibration is applied to the gloves of the participants when the robot collides with the environment.
Here we describe how the vibration intensity is determined.
We are inspired by a similar and very early study using appropriate "collision" signals to transmit variable frequency tactile feedback \cite{10.1007/3-540-44589-7_1}.
In a more previous study investigating vibrotactile approach, vibration intensity applied to users was proportional to the size of the virtual object being manipulated \cite{HapticFeedbackVibration}.
We adopt this approach where instead, vibration intensity is proportional to kinetic energy $K_E$ and object penetration $P$ of each finger segment in simulation.
These are then combined to give the final intensity. 

Kinetic energy $K_E$ of the virtual collision is formulated as:
\begin{equation}
K_{E}=\frac{1}{2} \cdot m \cdot v^{2} ,
\label{equation:KineticEnergy}
\end{equation}
where $m$ is the body mass and $v$ the velocity between the robot segment and the environment. 

We use the relative penetration of $P$ between the robot and the environment as a proxy for force.
Since our environment is simulated, we have access to the full state space of the environment.
Penetration can then be easily defined by the relative distance between the robot segment $v_{r}$ and virtual object $v_{o}$ and the distance between the centre and surface of the object $s_{o}$ as shown in \autoref{equation:ObjectPenetration}
\begin{equation}
P = \left | 1 - \frac{\left \| v_{r} - v_{o} \right \|}{ \frac{1}{2} \cdot s_{o}}    \right |  
\label{equation:ObjectPenetration}
\end{equation}

\begin{figure} [H]
\centering
  \includegraphics[width=\columnwidth]{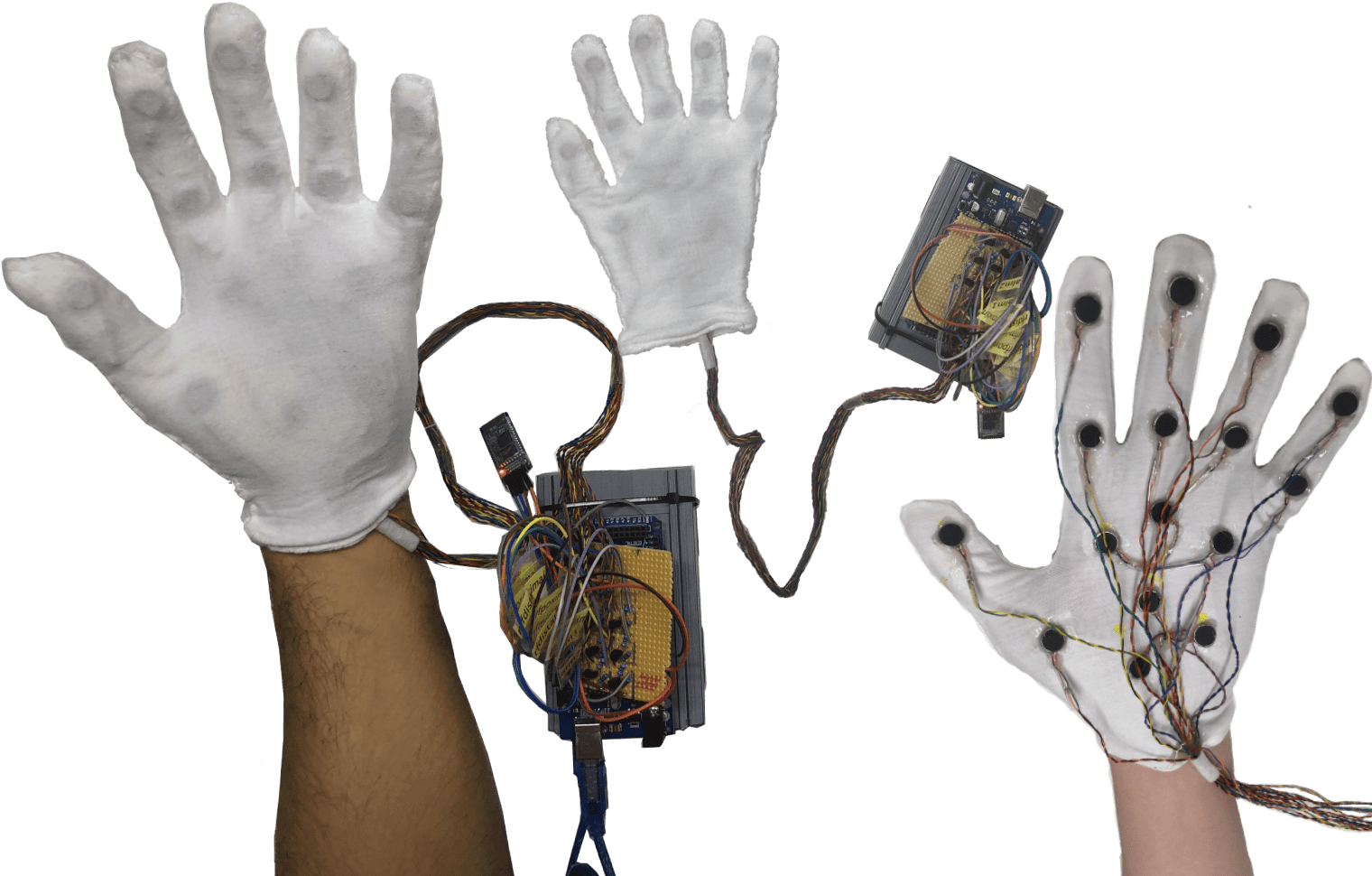}
        \vspace{-2.5mm}
  \caption{Haptic glove (left module shown) in its final and first iteration with it's electronics and motors exposed in the latter.}~\label{fig:HapticGlove}
\end{figure}

\autoref{equation:KineticEnergy} and \autoref{equation:ObjectPenetration} can then be combined to calculate total vibration intensity shown in \autoref{equation:FinalVibrationIntensity}
\begin{equation}
V=V_{min} + a \cdot \frac{\left (V_{max} - V_{min}  \right )}{K_{E_{max}}} \cdot K_{E} + b \cdot \frac{\left (V_{max} - V_{min}  \right )}{P_{max}}  \cdot \ P  
\label{equation:FinalVibrationIntensity}
\end{equation}
where $V$ is the final vibration intensity transmitted to the vibration motors, $V_{min}$ the minimum vibration intensity needed to distinguish vibrotactile stimulation when in contact.
This is set to 25\% based on a pilot study consisting of five participants. 
The second term calculates the vibration intensity based upon the kinetic energy exerted and is controlled by a constant $a$. $V_{max}$ is the maximum vibration intensity of the hardware, $K_{E_{max}}$ is the maximum calculated kinetic energy in Joules with a velocity limit of 7 $m/s$ set in the physics engine and $K_{E}$ is the current kinetic energy exerted to the object.
The kinetic energy is only applicable during the object acquisition and as masses are constant, it is only dependent upon the velocity of grabbing i.e. picking up.
The final term calculates the vibration intensity based upon the penetration of the robotic hand with the object and is controlled by a constant $b$. $P_{max}$ is the maximum penetration allowed which in our case is 100\% and finally, $P$ is the current penetration exerted to the object. \autoref{fig:HapticGlove} illustrates our haptic glove in addition to it\'s electronics, drive control board and motors exposed.

\subsection{Manipulation Tasks of Varying Complexity}
All tasks required the participants to pick up an object from a set starting point and place it to a designated random target location illustrated with a slight-transparent shape. 
We integrated three basic types of three-dimensional object shapes to not only introduce the inherent different complexities that come with such objects but also to be able to assess different grasping techniques \cite{MuJoCoHaptix,light_chappell_kyberd_2002}. While different shapes do indeed vary the task complexity, we also introduced different object sizes as well as placement distances. \\

\subsubsection{Task A - Cube Manipulation}
The first task included manipulating a cube shape. A cube was used, as it does not flip or roll and we can assess both its position and rotation accuracy. Grabbing techniques employed included Precision Grasping via Palmar Pinch \cite{light_chappell_kyberd_2002}. \\

\subsubsection{Task B - Cylinder Manipulation}
The second task included manipulating a cylinder shape. A cylinder can flip over and roll over a surface, making the task harder. We can also assess both the cylinder's position and rotation. Grabbing techniques employed included Precision Grasping via Palmar Pinch, as well as Cylindrical Grasping, also known as Power Grasp \cite{light_chappell_kyberd_2002}. \\

\subsubsection{Task C - Sphere Manipulation}
The third and final task was concerned with the manipulation of a sphere-shaped object. This was considered to be the hardest task due to the inherent ability of a sphere to roll over an even ideally horizontally placed surface if sufficient velocity would accumulate either from an inadequate precision velocity placement or release from a height offset. Grabbing techniques employed included Precision Grasping via Palmar Pinch as well as Spherical Grasping \cite{light_chappell_kyberd_2002}. \\

\subsubsection{Object Scale and Placement Distance}
\begin{figure} [H]
\centering
  \includegraphics[width=0.65\columnwidth]{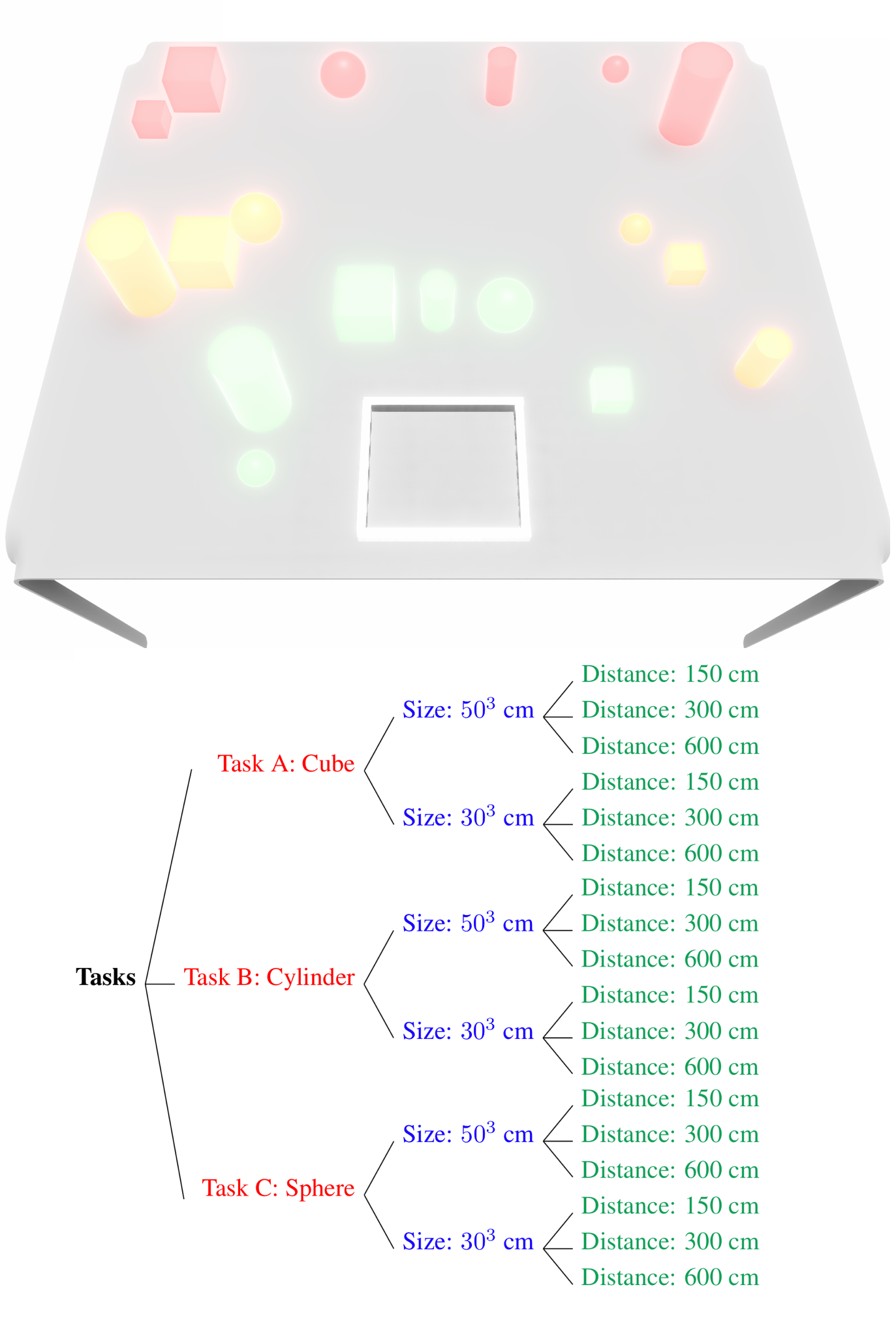}
        \vspace{-2.5mm}
  \caption{Image (Upper): All manipulation tasks illustrating the different three dimensional shapes, sizes as well as distances from 150, 300 and 600; green, yellow and red respectively. Tree (Bottom): All 18 tasks broken down in type of object shapes (red), sizes (blue) and distances (green). }~\label{fig:AllManipulationTasks}
\end{figure}

The aforementioned tasks are broken down into two sub-tasks assessing two object scales, large 50.0 x 50.0 x 50.0 (mm) (LxWxH) and small 30.0 x 30.0 x 30.0 (mm) (LxWxH). Furthermore, the aforementioned sub-tasks are broken down into sub-sub tasks assessing placement distances, defined as the absolute distance from the set starting point to a random target location with distances ranging from 150.0, 300.0 and 600.0 (mm), making it progressively more difficult. Taking into consideration our 8 interface conditions, two different object sizes, three different object shapes and distances, accounted to a total of 144 trials were conducted per participant. Across all participants, a total of 3600 trials were recorded. All of the manipulation tasks are visually depicted in \autoref{fig:AllManipulationTasks}. \\

\subsubsection{Task Progression and Succession}
Progression to the next task is achieved when there is an intersection between the actual object and target position, regardless of the accuracy. When an overlap is achieved, the target placement slightly glows and a two-second progression timer is initiated which only pauses when the object does not retain its position. This countdown only pauses when the object is no longer colliding with the target placement volume i.e. indicating that the object has either been moved or has not remained stationary. Task progression is also achieved if the countdown timer, which has been set to 30 seconds for all tasks, reaches zero, however, in that case, the task is considered a fail rather than a success. Finally, for all tasks, an invisible collision wall was implemented to avoid objects falling out of physical bounds rendering a retrieval impossible. 

\section{Evaluation}
To evaluate each interface across all manipulation tasks, we implemented both subjective and objective measurements, since immersion and perception are highly subjective and our tasks are objective.
We implemented both measurements to compensate for the inherent drawbacks of exclusively using questionnaires \cite{Slater:1994:DPV:2870936.2870938, 1219cba4b03b4d9e81481fd65a53a490}.
Measurements are summarized in \autoref{table:AssessmentOfConditions}.

\subsection{Subjective Measurements}
We first measured cognitive workload for each interface condition through the use of the multidimensional assessment tool questionnaire NASA-Task Load Index, simply known as NASA-TLX \cite{hart_staveland_1988}. Incorporating six sub-scales including mental demand, physical demand, temporal demand, effort, frustration, and performance.

In addition, we assessed overall system usability, through the use of the System Usability Scale questionnaire, just known as SUS \cite{brooke1996sus}. Consisting of ten in total questions on a 5-point Likert scale, which range from "strongly disagree" to "strongly agree", evaluating system complexity, consistency and cumbersomeness.

\subsection{Objective Measurements}
Overall task performance was measured by first comparing the total proportion of successful task completion, defined as placing the object to the target location within a time-countdown window of 30 seconds for each task regardless of accuracy, however, a minimum overlap with the target volume was required.

Time-based metrics were also incorporated, specifically placement and completion time to assess how fast performing each interface was. Placement time was defined as the time it took users to pick up the object and place it to the target location with potential accuracy corrections afterwards not being assessed, strictly the time stamp of the object and the target volume being in their very first collision. Completion time, on the other hand, was defined as the overall time it took users to complete successfully a task. 

In addition to time, spatial-based metrics were also implemented to assess the accuracy of placing objects and how each interface may affect these, which is vital in remotely piloted systems concerned with fine manipulation. Target distance error was considered at the end of each task and defined as the distance between the center of the object and the target location, with higher values indicating worse performance. In addition, position accuracy was calculated by averaging all three axes from the euclidean space center of the object and the target (X,Y,Z) in one final percentage value. Orientation accuracy was similarly calculated but dependent upon the three-dimensional shape. For the cube and cylinder, a modulo operation of 45 and 90 degrees was performed respectively. Assessing the orientation of the sphere was disregarded due to its inherent shape.

\begin{table}[H] \centering
\begin{footnotesize}
\begin{tabular}{lrrr} \toprule
 \textbf{Measurement} & & Type & Metric \\ \midrule

\rowcolor{Gray}\begin{tabular}[c]{@{}l@{}}Cognitive Workload\end{tabular}   & &
Subjective
&
Questionnaire [\textit{Likert Scale}]
\\  \addlinespace[0.1cm]
\begin{tabular}[c]{@{}l@{}}System Usability\end{tabular}   & &
Subjective
&
Questionnaire [\textit{Likert Scale}]
\\  \addlinespace[0.1cm]
\rowcolor{Gray}\begin{tabular}[c]{@{}l@{}}Task Succession\end{tabular}   & &
Objective
&
Percentage [\textit{\%}]
\\ \addlinespace[0.1cm]
\begin{tabular}[c]{@{}l@{}}Placement Time\end{tabular}   & &
Objective
&
Seconds [\textit{s}]
\\  \addlinespace[0.1cm]
\rowcolor{Gray}\begin{tabular}[c]{@{}l@{}}Completion Time\end{tabular}   & &
Objective
&
Seconds [\textit{s}]
\\  \addlinespace[0.1cm]
\begin{tabular}[c]{@{}l@{}}Target Error\end{tabular}   & &
Objective
&
Meters [\textit{m}]
\\  \addlinespace[0.1cm]
\rowcolor{Gray}\begin{tabular}[c]{@{}l@{}}Position Accuracy\end{tabular}   & &
Objective
&
Percentage [\textit{\%}]
\\  \addlinespace[0.1cm]
\begin{tabular}[c]{@{}l@{}}Rotation Accuracy\end{tabular}   & &
Objective
&
Percentage [\textit{\%}]
\\  
\bottomrule
\end{tabular}
\end{footnotesize}
\caption{Summary of both objective and subjective measurements.}
\label{table:AssessmentOfConditions}
\end{table}

\subsection{Procedure}
Prior to commencing the experiment, participants were briefed on the purpose of the experiment, gave formal written consent and were handed out the NASA-TLX as well as the SUS questionnaires to allow acquaintance with the scales. Once users got familiar with the questionnaires, their interpupillary distance (IPD), was measured for the VRHMD and they were allowed for 10 minutes to get acquainted with the simulation environment. 
During this acclimatization procedure, the participants were able to familiarize themselves with the keyboard controls and the technologies implemented in the actual experiment, but not with the actual tasks. Furthermore, due to having eight different interface conditions, we also randomized the order of these multi-modal interfaces for each participant, to counterbalance potential acclimatization or task adaption.

\section{Results}
\subsection{Analyses Techniques and Methods}
To analyze our results, where parametric and without normality violation, via a Shapiro-Wilk Test, a repeated-measures analysis of variance was used (RM-ANOVA) in addition to post-hoc analysis for pairwise comparison of the eight different interface conditions. Where sphericity was not met, via Maulchy's Test, a Greenhouse-Geisser correction was used to account for the violation and correct the degrees of freedom assuming a $\epsilon < 0.75$, otherwise a Huynh-Feld correction was used \cite{abdi2010greenhouse}. 

For non-parametric data, specifically for ordinal data i.e. likert scales, an Aligned-Rank Transform (ART) \cite{Wobbrock:2011:ART:1978942.1978963} was used to allow the use of parametric tests i.e. RM-ANOVA. For non-parametric continuous data, a Friedman's test, similar to the RM-ANOVA, was used to test for significance across the eight interface conditions \cite{doi:10.1080/01621459.1937.10503522}, and Wilcoxon signed-rank tests for post-hoc analysis for the pairwise comparison across the interface conditions. 

Samples that were classified as a Bernoulli distribution, the proportion of successful completion, a two times standard deviation from the mean was considered significant (95\% CI) i.e. empirical rule \cite{10.2307/2684253}. Hereinafter, for all reported results, the significance levels are: * $p<0.05$, ** $p<0.01$, *** $p<0.001$ and n.s not significant. 

Finally, in the Appendices, we summarize the overall results of each interface conditions across all measurements, thus giving new evidence to the hypothesized and untested effectiveness of each interface condition suggested by Sigrist et al. \cite{Sigrist2013}.

\subsection{Subjective Results}
\subsubsection{Perceived Workload}
\begin{figure}[H]
\centering
  \includegraphics[width=\columnwidth]{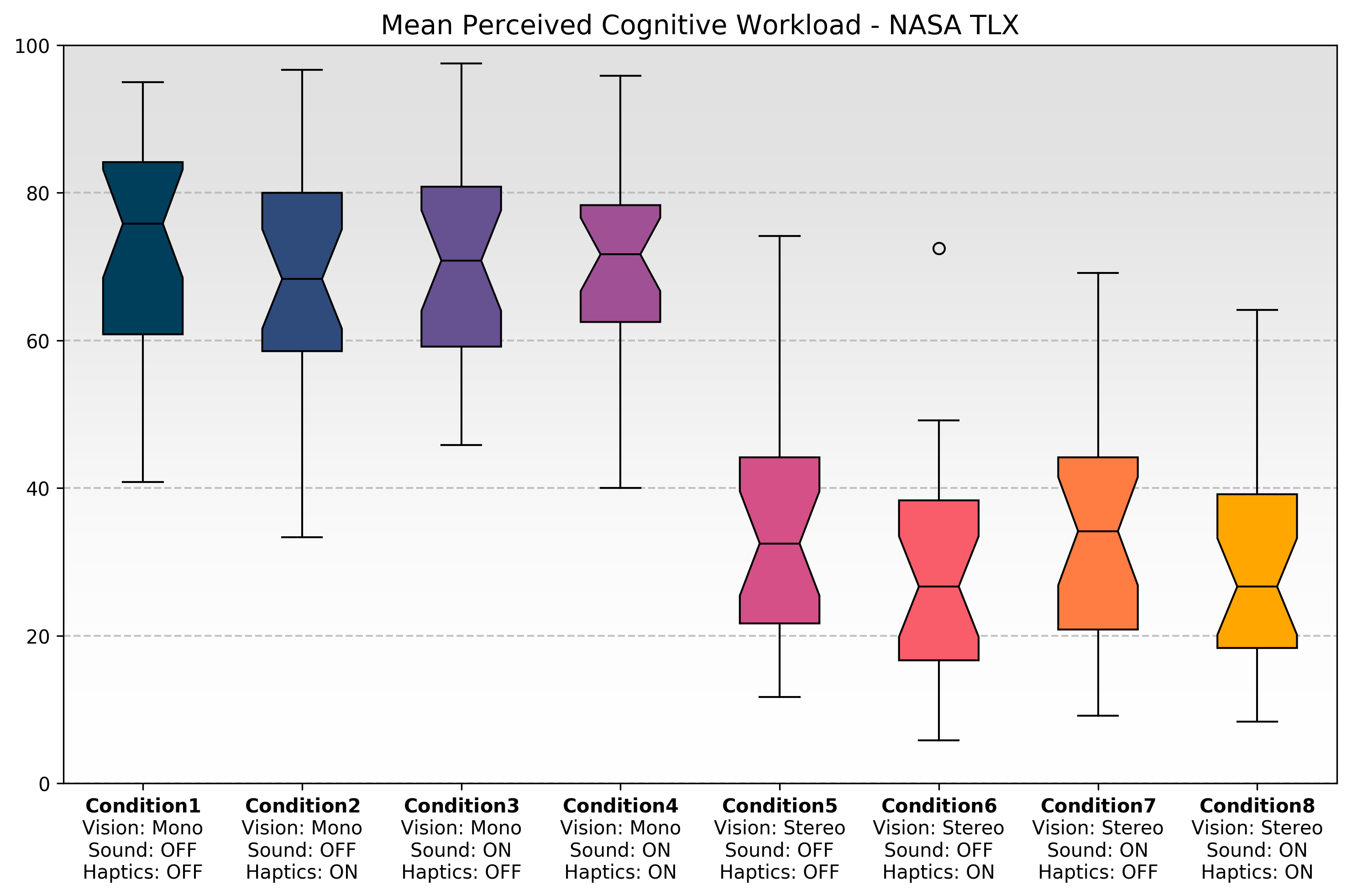}
        \vspace{-7.5mm}
\caption{Box plot illustration across all eight interface conditions of the mean perceived workload, with higher scoring equal to worse performance. Dots represent outliers.}~\label{fig:NasaTLX}
\end{figure}

For the perceived workload, an ART was used to allow the use of parametric tests on ordinal data. A one way RM-ANOVA with a Greenhouse-Geisser correction ($\epsilon=0.342$) was used, yielding a highly significant difference across all eight interface conditions, ($F(2.393,57.437)=70.473$,$p<0.001$,$\eta_{p}^2=0.746$). 
Mean responses for perceived workload demand are shown in \autoref{fig:NasaTLX} and \autoref{table:SummaySubjectiveMeasurements}. Post-hoc analysis showed partial support of hypothesis H1, specifically (a) that conditions incorporating monocular vision with the display monitor, C1,2,3 \& 4, accounted to significantly higher perceived workload ($p<0.001$) than  stereoscopic vision with the VRHMD, C5, C6, C7 \& C8. Furthermore, (b) conditions incorporating somatosensory feedback only when paired with stereoscopic feedback. C6 \& 8, showed significantly lower perceived workload ($p<0.05$) than those who do not: C5 \& 7 and when paired with monocular feedback, marginally lower workload was observed with somatosensory C2 ($p=0.056$), than only monocular C1. Finally, (c) conditions with audition only did not contribute to an observable difference in workload ($p>0.05$).

\subsubsection{Interface Usability}
\begin{figure}[H]
\centering
  \includegraphics[width=\columnwidth]{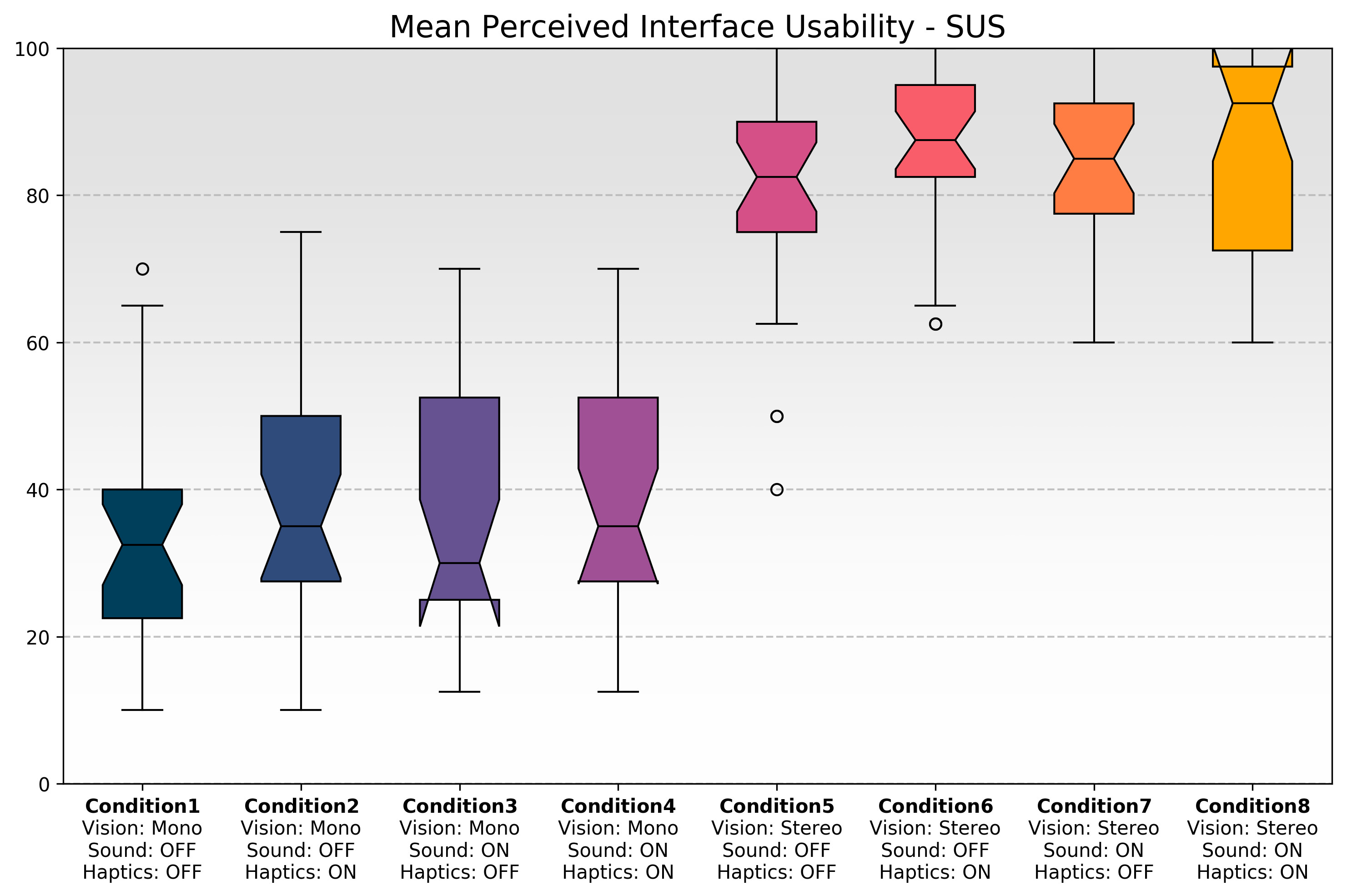}
        \vspace{-7.5mm}
\caption{Box plot illustration across all eight interface conditions of the mean interface usability, with higher scoring equal to better performance. Dots represent outliers.}~\label{fig:SUS}
\end{figure}

For the perceived system usability, an ART was used to allow the use of parametric tests on ordinal data. A one way RM-ANOVA with a Greenhouse-Geisser correction ($\epsilon=0.561$) was used, yielding a highly significant difference across all eight interface conditions, ($F(3.930,94.310)=97.064$,$p<0.001$,$\eta_{p}^2=0.802$). 
Average responses for interface usability are shown in \autoref{fig:SUS} and \autoref{table:SummaySubjectiveMeasurements}. Post-hoc analysis revealed that the same trend holds true for the system usability, as with the cognitive workload. Specifically, we again found partial support of our H2 hypothesis with (a) stereoscopic vision with the VRHMD C5,C6,C7 and C8 accounting to significantly higher interface usability than monocular vision with the display monitor C1,C2,C3 and C4, ($p<0.001$), (b) somatosensory feedback further increasing overall usability however again only when paired with stereoscopic visual feedback C6 and C8 ($p<0.05$), and finally (c) auditory feedback by itself making no significant difference ($p>0.05$).

\begin{table}[H]
\centering
\begin{scriptsize}
  \begin{tabular}{lccccccc}
    \toprule
      \multicolumn{3}{l}{\textbf{Subjective Measurements}} &  &
      \multicolumn{2}{l}{\textbf{NASA-TLX}} &
      \multicolumn{2}{l}{\textbf{SUS}} \\
      \cmidrule(lr){5-6}
      \cmidrule(lr){7-8}
      & \textit{Vision} & \textit{Audio} & \textit{Haptic} & {\textit{Med.}} & {\textit{Std. D.}} & {\textit{Med.}} & {\textit{Std. D.}} \\ 
      \midrule
       \rowcolor{Gray}C1 & Monitor & Off & Off & 75.83 & $\pm$13.82  & 32.50 & $\pm$14.34  \\
       C2 & Monitor & Off & On & 68.33 &  $\pm$16.22 & 35.00 & $\pm$17.15 \\
       \rowcolor{Gray}C3 & Monitor & On & Off & 70.83 & $\pm$15.00  &  30.00 & $\pm$17.76 \\   
       C4 & Monitor & On & On & 71.66 & $\pm$14.93 & 35.00  & $\pm$15.63  \\ 
       \rowcolor{Gray}C5 & VRHMD & Off & Off & 32.50  & $\pm$16.73 & 82.50  & $\pm$15.82 \\
       C6 & VRHMD & Off & On & 26.66 & $\pm$15.91 & 87.50  & $\pm$11.33 \\
       \rowcolor{Gray}C7 & VRHMD & On & Off & 34.16  & $\pm$15.58 & 85.00 & $\pm$11.38 \\
       C8 & VRHMD & On & On & 26.66  & $\pm$14.52 & 92.50 & $\pm$13.20 \\
    \bottomrule
  \end{tabular}
\end{scriptsize}
  \caption{Summary of all subjective results, reporting median and standard deviation across all eight interface conditions.}
  \label{table:SummaySubjectiveMeasurements}
\end{table}

\subsection{Objective Results}
\subsubsection{Error Rate}
\begin{figure}[H]
\centering
  \includegraphics[width=\columnwidth]{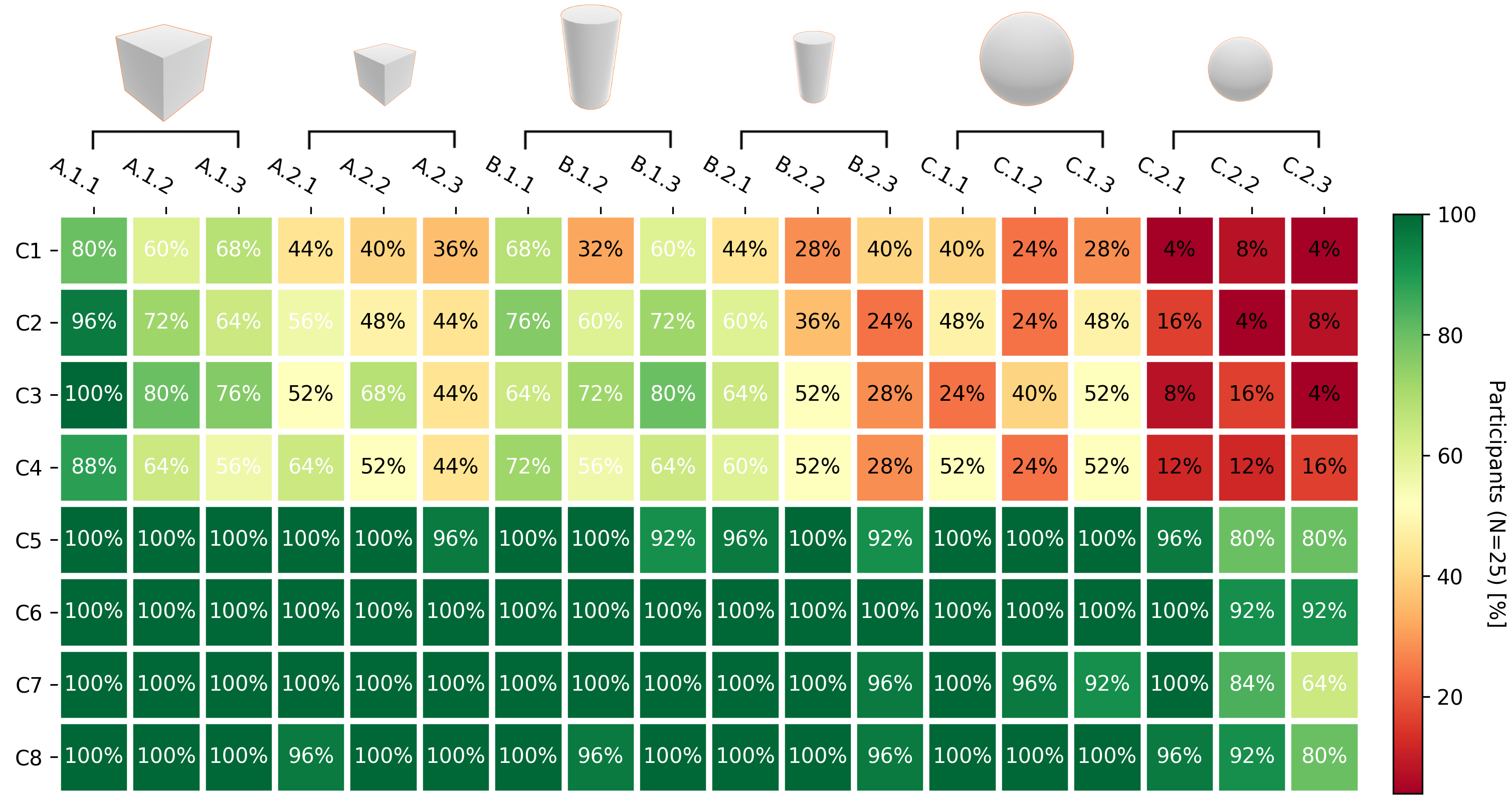}
        \vspace{-2.5mm}
  \caption{Heat-map illustrating the proportion of task success rate going from lower to higher complexity, horizontal axis A.1.1 (left) to C.2.3 (right), across all the interface conditions C1 to C8, vertical axis.}~\label{fig:HeatMapPercentage}
\end{figure}

First, we analyzed the total proportion of successful task completion (\%), across all interface conditions. Our sample was classified as a Bernoulli distribution and a two-times standard deviation from the mean, three-sigma rule, was used to test for significance. Results show that, interface conditions incorporating stereoscopic vision with the VRHMD (C5,C6,C7,C8) accounted to a significant observable difference, ($p<0.05$), in mean success rates 96.22\% (SD=4.73\%), 99.11\% (SD=2.62\%), 96.22\% (SD=5.94\%), 97.55\% (SD=4.26\%) respectively compared to the monocular display monitor (C1,C2,C3,C4) with rates 39.33\% (SD=21.69\%), 47.55\% (SD=18.77\%), 51.33\% (SD=23.63\%), 48.22\% (SD=20.26\%) respectively. No significant differences were observed between conditions incorporating haptic or auditory feedback ($p>0.05$). Results are depicted in \autoref{fig:HeatMapPercentage}.

\subsubsection{Placement and Completion Time}
\begin{figure*}
\centering
\subfigure{\label{fig:aa}\includegraphics[width=4.25cm]{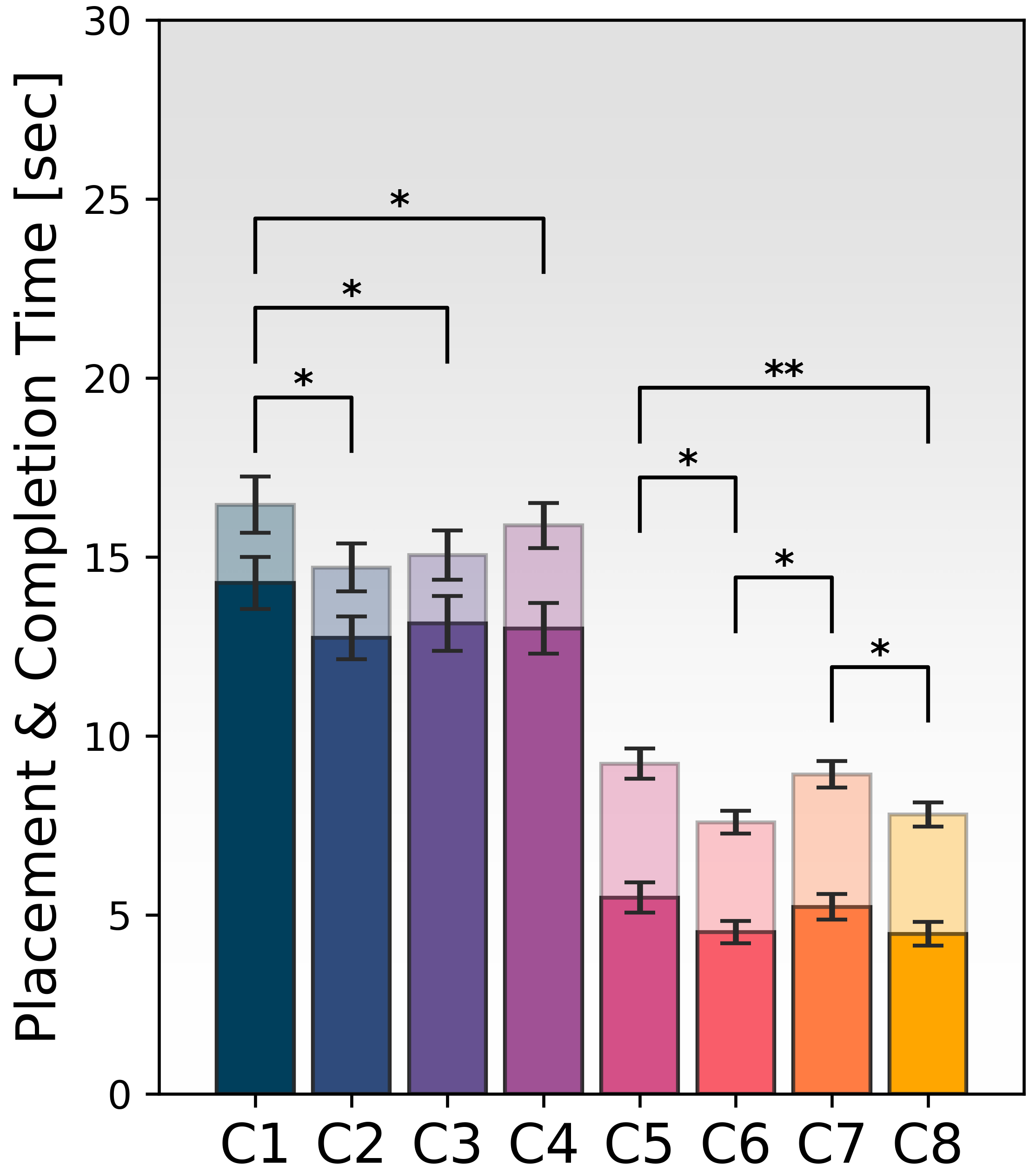}}
\subfigure{\label{fig:ab}\includegraphics[width=4.375cm]{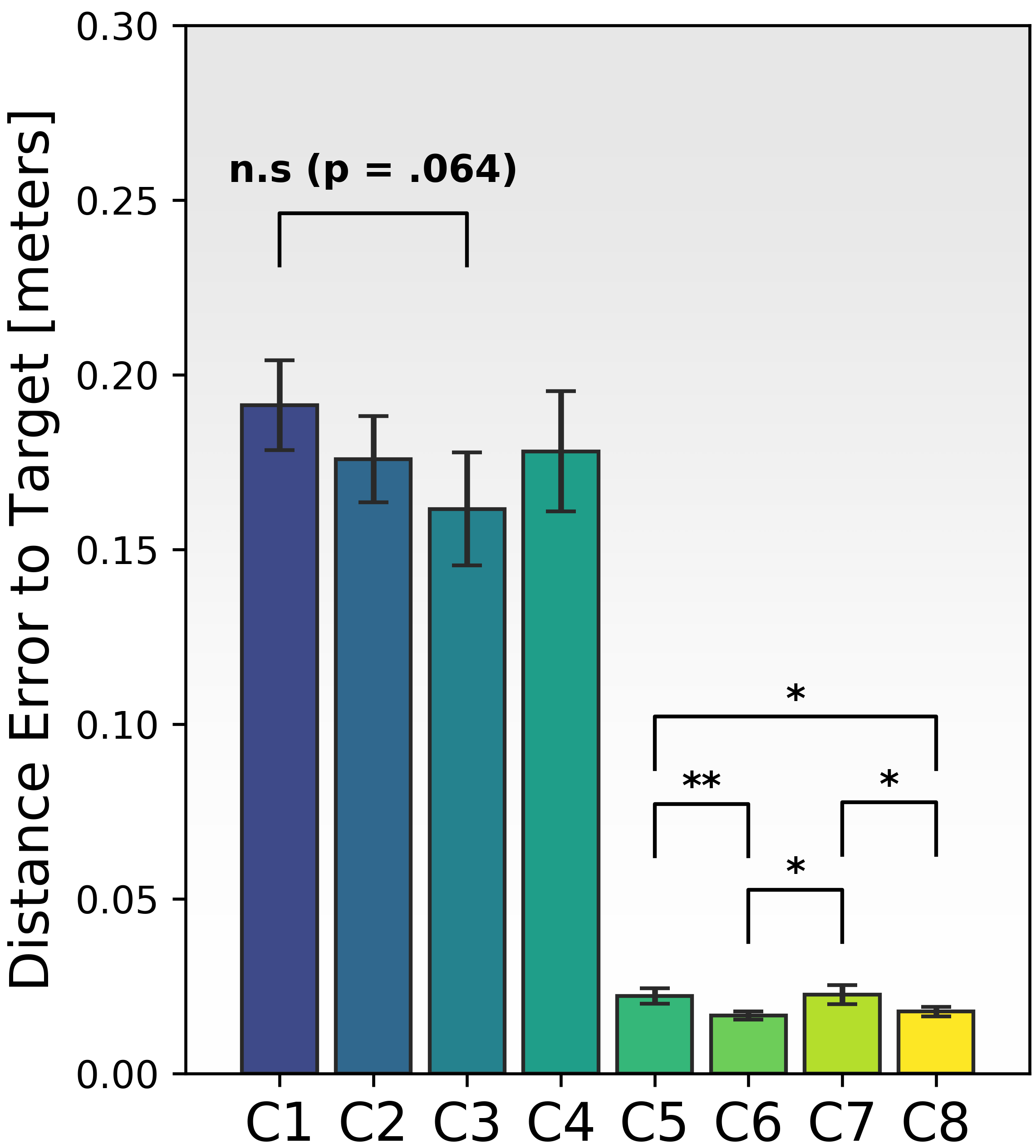}}
\subfigure{\label{fig:ac}\includegraphics[width=4.35cm]{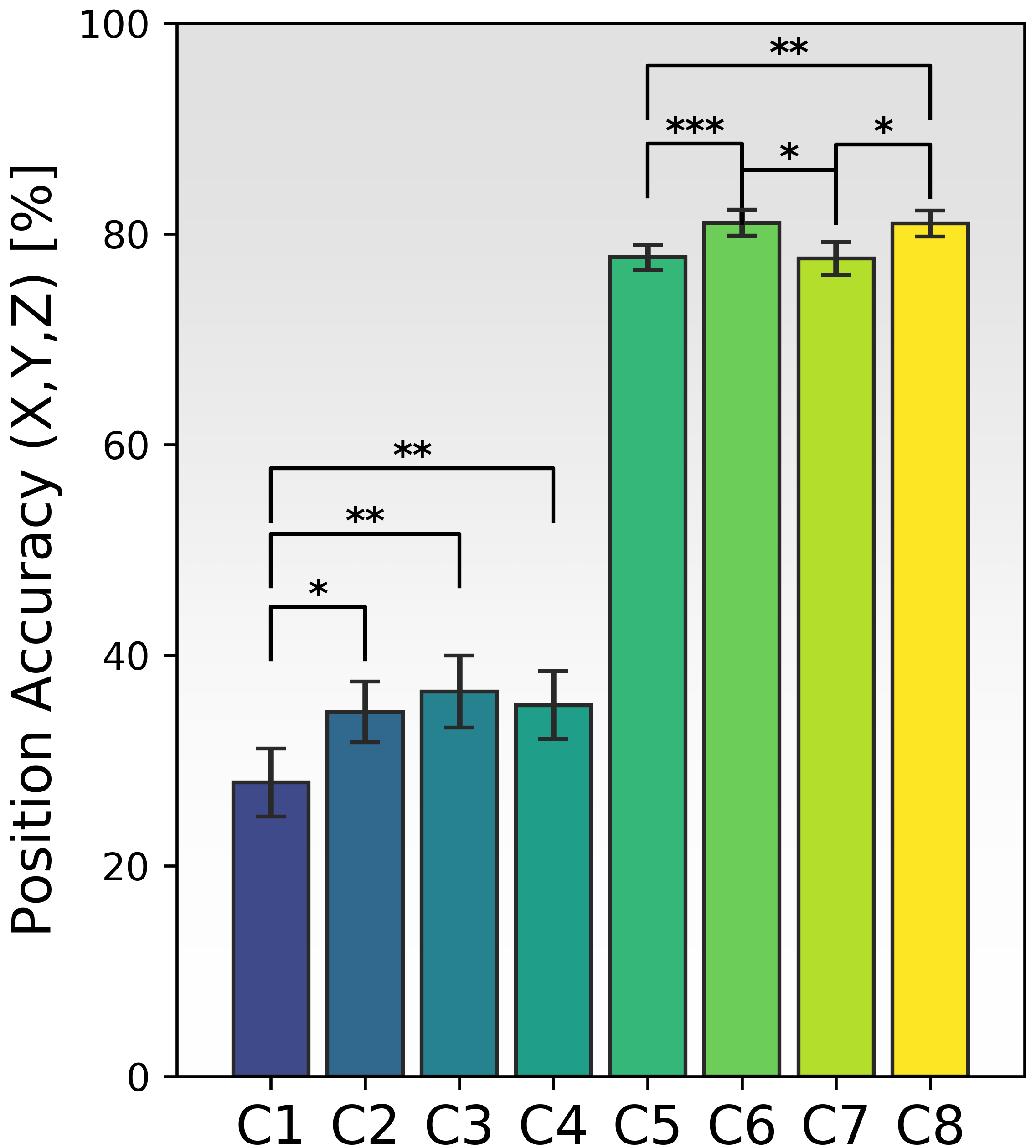}}
\subfigure{\label{fig:ad}\includegraphics[width=4.35cm]{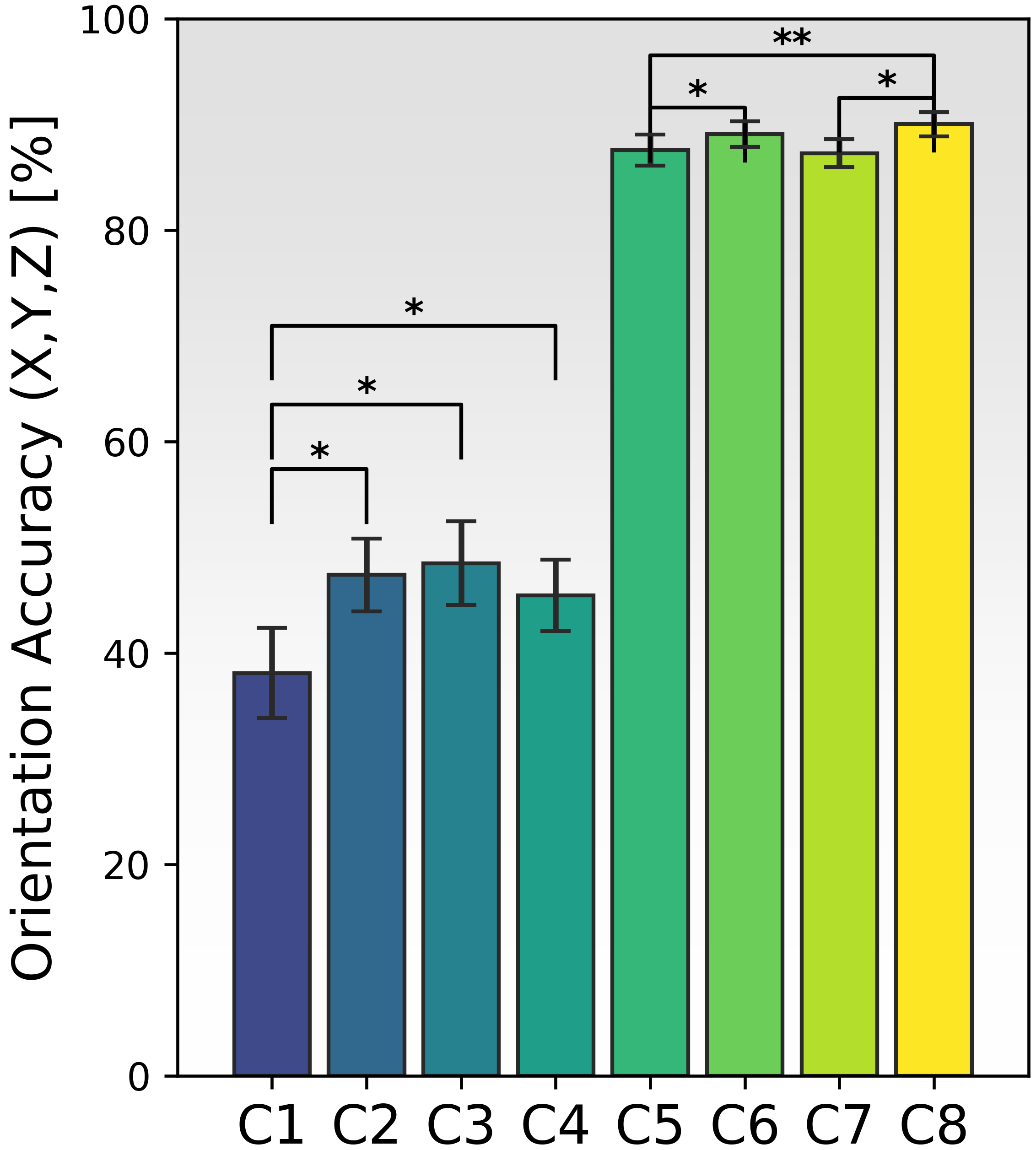}}
      \vspace{2.5mm}
\caption{Objective measurements represented in a bar graph in addition to standard error. From left to right, time-based metrics mean placement (opaque) and completion time (slightly transparent). Followed by spatial based metrics, specifically distance error, position and rotation accuracy.}~\label{fig:SummaryObjectiveMeasurementsOnly}
\end{figure*}

For time-based metrics we considered only the successful instances. Transforming our data in a non-parametric state, Shapiro-Wilk Test for normality yielded ($p<0.05$) in both instances. Friedman's test was thus used, yielding a significant difference in mean placement as well as completion time across the eight interface conditions ($\chi^2(2)=129.093$, $p<0.001$) and ($\chi^2(2)=131.093$, $p<0.001$) respectively. Placement and completion times are shown in \autoref{table:SummayObjectiveMeasurements} and \autoref{fig:SummaryObjectiveMeasurementsOnly}. Post-hoc analysis using the Wilcoxon Signed-Rank tests showed partial support of our H3 hypothesis, specifically (a) stereoscopic visual feedback with the VRHMD, C5,C6,C7 and C8 accounted to highly significantly less placement and completion time than with the monocular display monitor ($p<0.001$) C1, C2 C3 and C4, followed by (b) somatosensory feedback contributing additionally to significantly lesser placement and completion, however, only when paired with the VRHMD, C6 and C8 ($p<0.05)$. Auditory feedback (c), did not contribute to an observable difference across all conditions ($p>0.05$). 

\begin{table*}
\centering
\begin{scriptsize}
  \begin{tabular}{lccccccccccccc}
    \toprule
      \multicolumn{3}{l}{\textbf{Objective Measurements}} &  &
      \multicolumn{2}{l}{\textbf{Placement Time [s]}} &
      \multicolumn{2}{l}{\textbf{Completion Time [s]}} &
      \multicolumn{2}{l}{\textbf{Distance Error [cm] }} &
      \multicolumn{2}{l}{\textbf{Pos Accuracy (XYZ) [\%]}} &
      \multicolumn{2}{l}{\textbf{Rot Accuracy (XYZ) [\%]}} \\
      \cmidrule(lr){5-6}
      \cmidrule(lr){7-8}
      \cmidrule(lr){9-10}
      \cmidrule(lr){11-12}
      \cmidrule(lr){13-14}
      & \textit{Vision} & \textit{Audio} & \textit{Haptics} & {\textit{Mean}} & {\textit{Std. D.}} & {\textit{Mean}} & {\textit{Std. D.}} & {\textit{Mean}} & {\textit{Std. D.}} & {\textit{Mean}} & {\textit{Std. D.}} & {\textit{Mean}} & {\textit{Std. D.}}  \\ 
      \midrule
       \rowcolor{Gray} C1 & Monitor & Off & Off & 14.27 & $\pm$3.65 & 16.45 & $\pm$3.91 & 19.12 & $\pm$6.42 & 27.89\% & $\pm$16.06 & 38.09\% & $\pm$21.35 \\
       C2 & Monitor & Off & On & 12.73 & $\pm$2.98 & 14.70 & $\pm$3.32 & 17.58 & $\pm$6.17 & 34.59\% & $\pm$14.30 & 47.37\% & $\pm$17.27 \\
       \rowcolor{Gray} C3 & Monitor & On & Off  & 13.14 & $\pm$3.83 & 15.05 & $\pm$3.42 & 16.15 & $\pm$8.06 & 36.50\% & $\pm$17.09 & 48.48\% & $\pm$19.86 \\
       C4 & Monitor & On & On  & 13.00 & $\pm$3.55 & 15.88 & $\pm$3.15 & 17.80 & $\pm$8.61 & 35.22\% & $\pm$16.11 & 45.44\% & $\pm$16.97 \\
       \rowcolor{Gray} C5 & VRHMD & Off & Off & 5.48 & $\pm$2.10 & 9.22 & $\pm$2.10 & 2.21 & $\pm$1.08 & 77.75\% & $\pm$5.93 & 87.55\% & $\pm$7.30 \\
       C6 & VRHMD & Off & On & 4.51 & $\pm$1.57 & 7.58 & $\pm$1.60 & 1.65 & $\pm$0.58 & 81.04\% & $\pm$6.08 & 89.08\% & $\pm$6.07 \\
       \rowcolor{Gray} C7 & VRHMD & On & Off & 5.22 & $\pm$1.77 & 8.92 & $\pm$1.85 & 2.26 & $\pm$1.37 & 77.65\% & $\pm$7.75 & 87.27\% & $\pm$6.65 \\
       C8 & VRHMD & On & On & 4.47 & $\pm$1.64 & 7.80 & $\pm$1.71 & 1.76 & $\pm$0.70 & 80.97\% & $\pm$6.16 & 90.01\% & $\pm$5.67 \\
    \bottomrule
  \end{tabular}
\end{scriptsize}
  \caption{Summary of objective results including time-based and spatial-based metrics with mean and standard deviation across all interfaces.}
  \label{table:SummayObjectiveMeasurements}
\end{table*}

\subsubsection{Distance Error}
For distance-error to target, data was normally distributed, Shapiro-Wilk ($p>0.05$). As such, a one way RM-ANOVA with a Greenhouse-Geisser correction was used ($\epsilon=0.381$), yielding a highly statistical significance across the eight interface conditions, ($F(2.664,63.950)=90.463$,$p<0.001$,$\eta_{p}^2=0.790$). Distance error across all interfaces is shown in
\autoref{table:SummayObjectiveMeasurements} and visually represented in \autoref{fig:SummaryObjectiveMeasurementsOnly}. Post-hoc analysis revealed a partial support of our H4 hypothesis, specifically (a) conditions incorporating stereoscopic vision with the VRHMD, C5, C6, C7,C8 accounted to significantly lower distance error ($p<0.001$), compared to conditions incorporating monocular vision with the display monitor, C1,C2,C3,C4. Furthermore, (b) conditions incorporating somatosensory feedback, however only when paired with stereoscopic visual feedback, C6, C8, showed further significantly lower target error and by extent higher accuracy to the target placement ($p<0.05$), than conditions without haptic feedback C5, C7 respectively. Finally, (c) conditions incorporating only auditory stimulation did not contribute to an observable difference in spatial accuracy than those without ($p>0.05$).

\subsubsection{Position and Orientation Accuracy}
Regarding spatial accuracy, specifically position and orientation accuracy, Shapiro-Wilk Test in both instances yielded ($p>0.05$) thus signifying normally distributed data. As such a one way RM-ANOVA with a Greenhouse-Geisser correction ($\epsilon=0.476$) and ($\epsilon=0.448$) respectively, yielded in both instances a highly significant difference ($F(3.332,79.962)=174.488$,$p<0.001$,$\eta_{p}^2=0.879$) and ($F(3.139,75.334)=109.280$,$p<0.001$,$\eta_{p}^2=0.820$) respectively. Position and orientation accuracy are shown in \autoref{table:SummayObjectiveMeasurements} and visually represented in \autoref{fig:SummaryObjectiveMeasurementsOnly}. Post-hoc analysis revealed a full support of our H5 hypothesis, specifically (a) conditions incorporating stereoscopic vision with the VRHMD, C5, C6, C7,C8 accounted to significantly higher spatial accuracy both in position and orientation ($p<0.001$), than conditions incorporating monocular vision with the display monitor, C1,C2,C3,C4. Furthermore, (b) conditions incorporating somatosensory feedback C2, C4 and C6, C8, showed further significantly higher spatial accuracy ($p<0.05$), than those who do not C1, C3 and C5, C7 respectively. Finally, (c) conditions incorporating only auditory stimulation C3 did also cause a greater increase in spatial accuracy than those without (C1) ($p<0.05$). Our findings here suggest that spatial accuracy increases significantly when stereo vision is used and furthermore when paired with either sound or somatosensory or even both, than just only relying on vision.

\begin{figure}[H]
\centering
\subfigure{\label{fig:a1}\includegraphics[width=0.49\columnwidth]{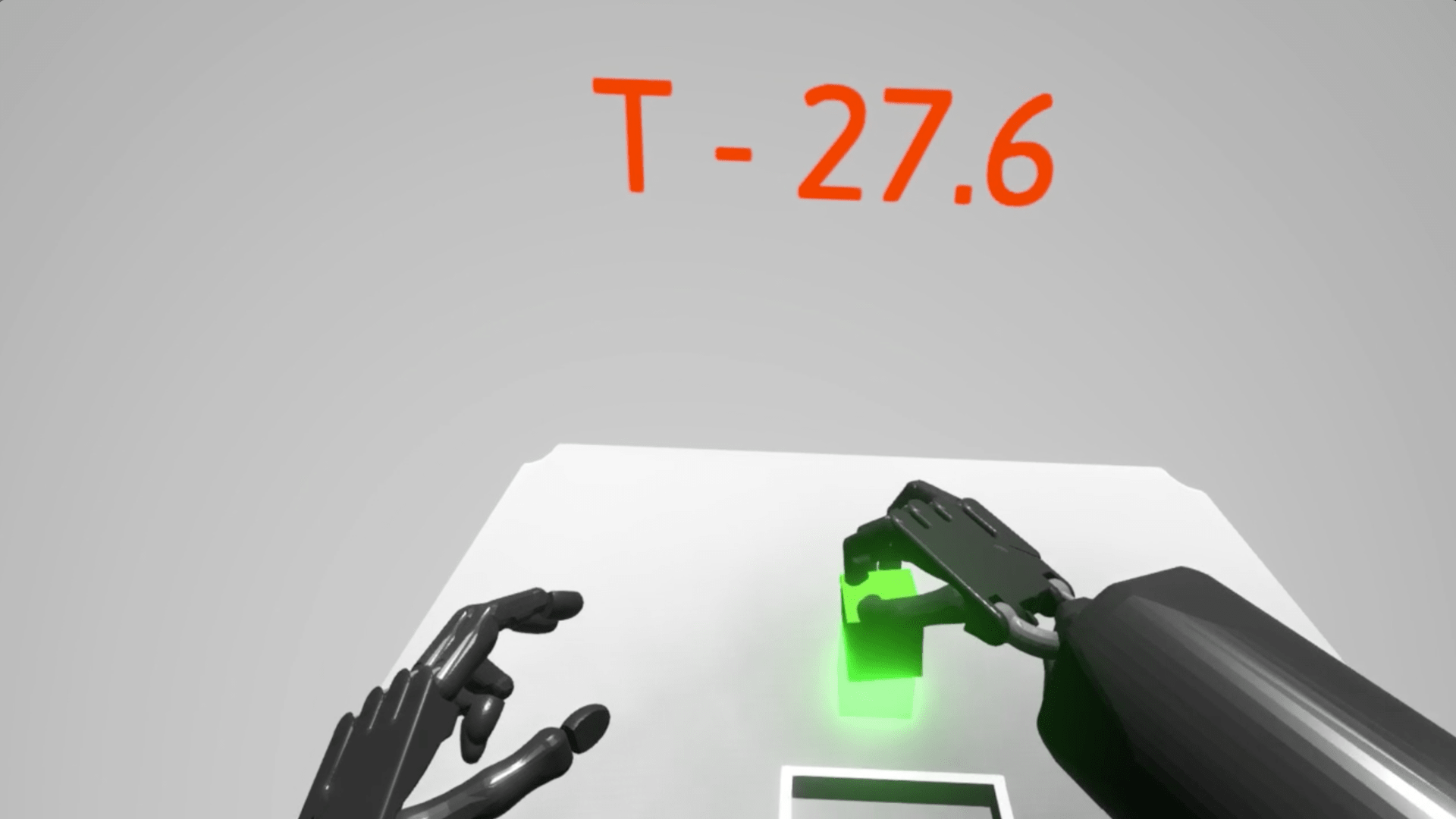}}
\subfigure{\label{fig:a2}\includegraphics[width=0.49\columnwidth]{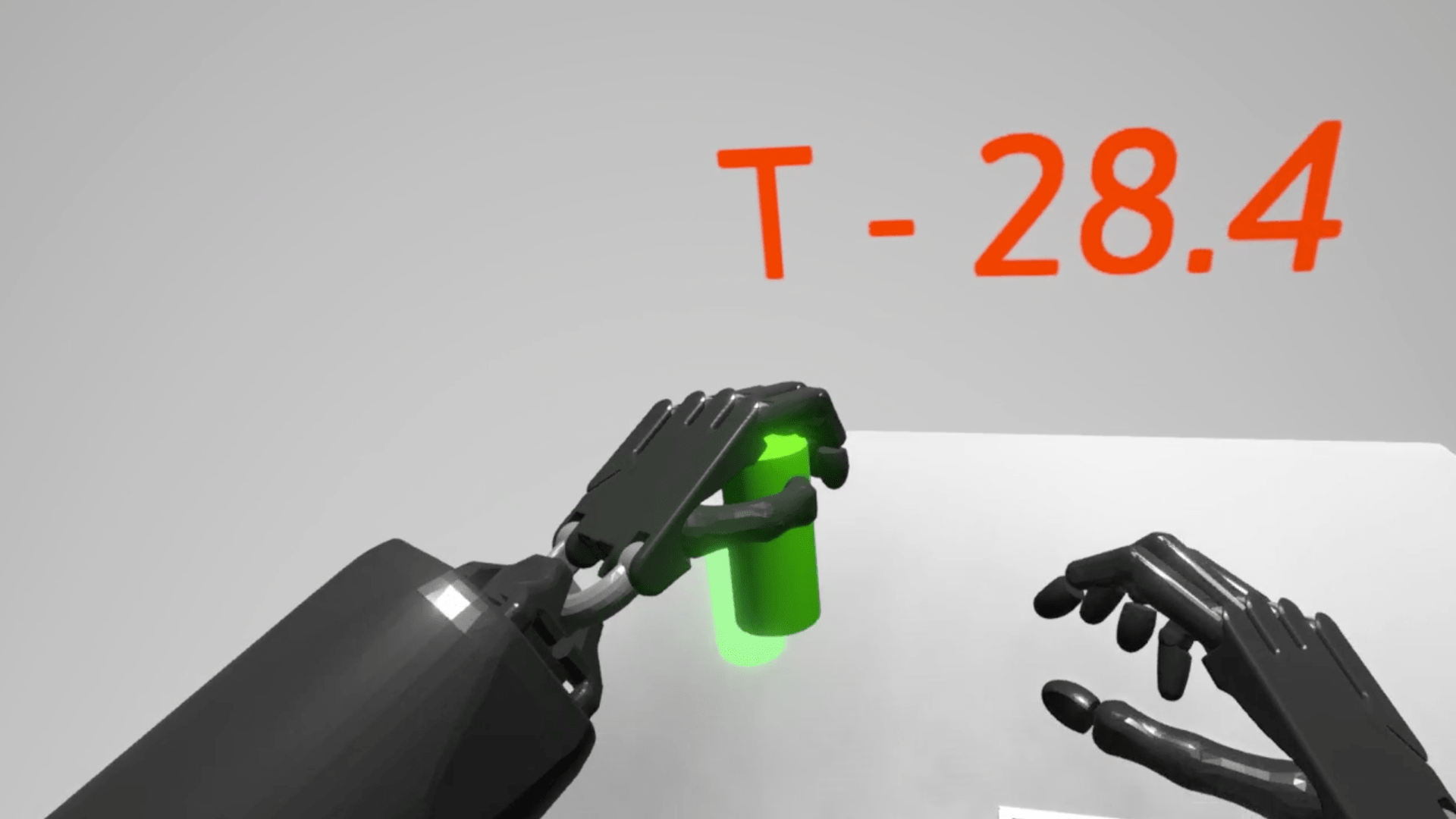}}
\subfigure{\label{fig:a3}\includegraphics[width=0.49\columnwidth]{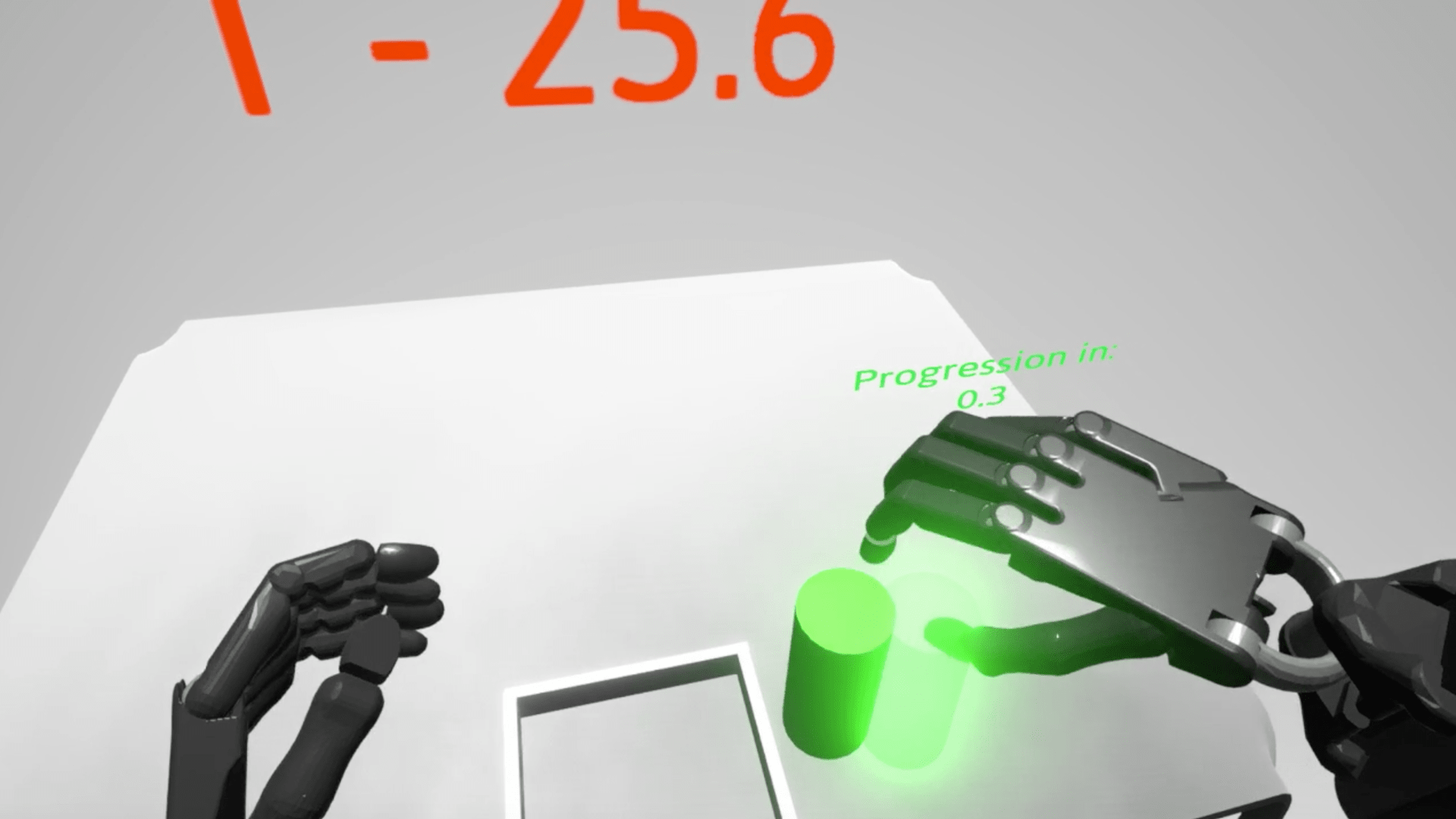}}
\subfigure{\label{fig:a4}\includegraphics[width=0.49\columnwidth]{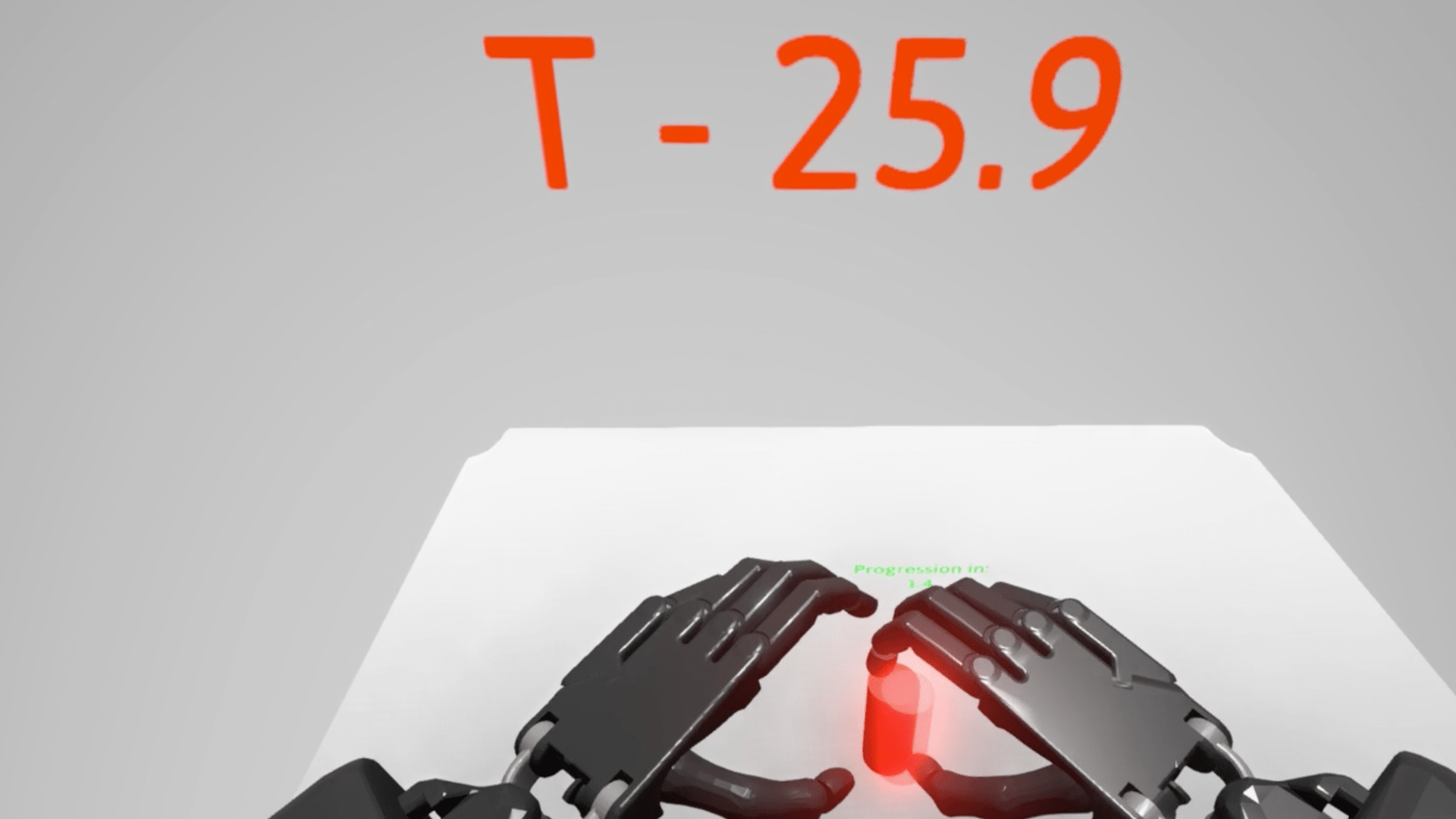}}
\subfigure{\label{fig:a5}\includegraphics[width=0.49\columnwidth]{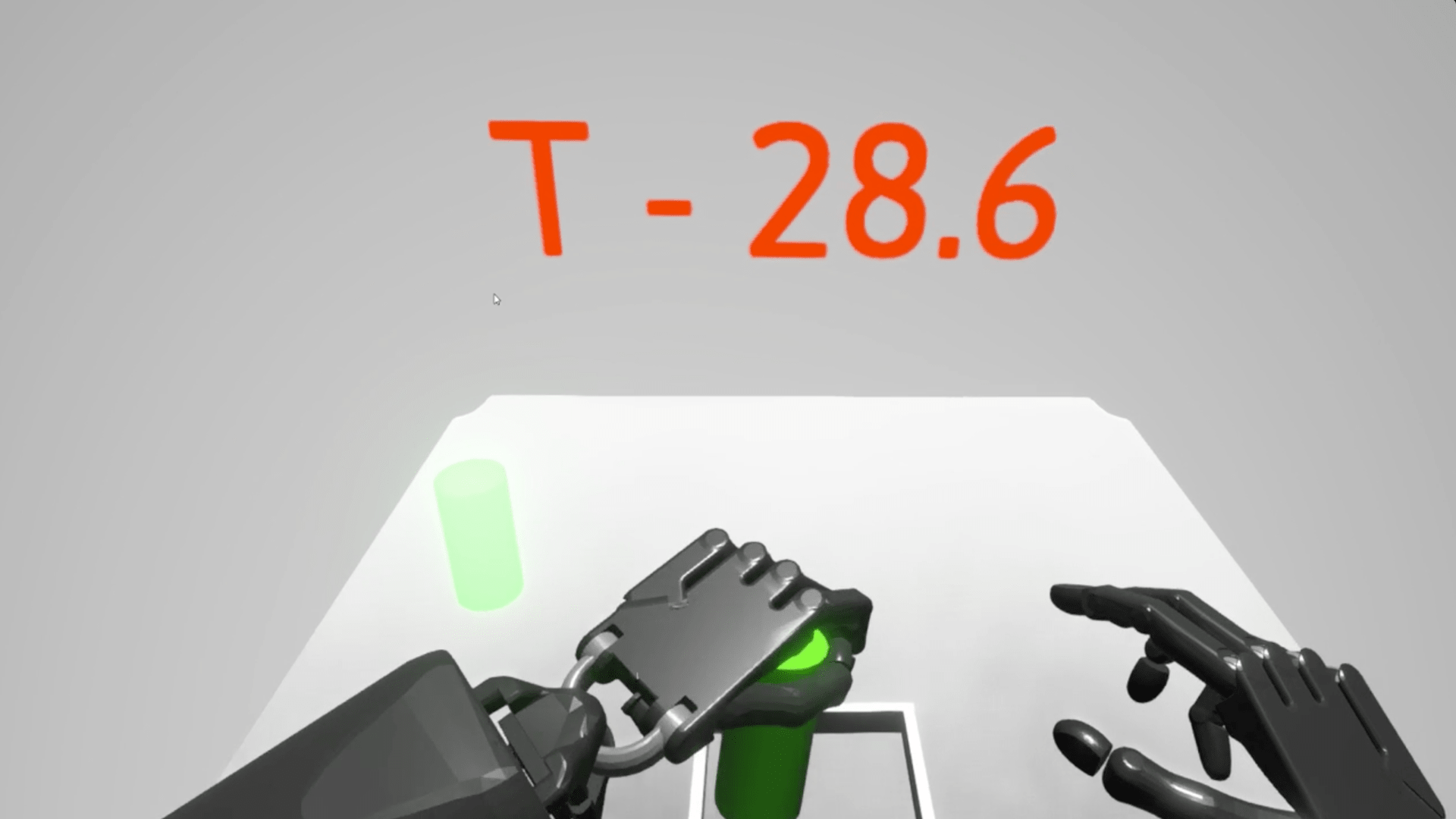}}
\subfigure{\label{fig:a6}\includegraphics[width=0.49\columnwidth]{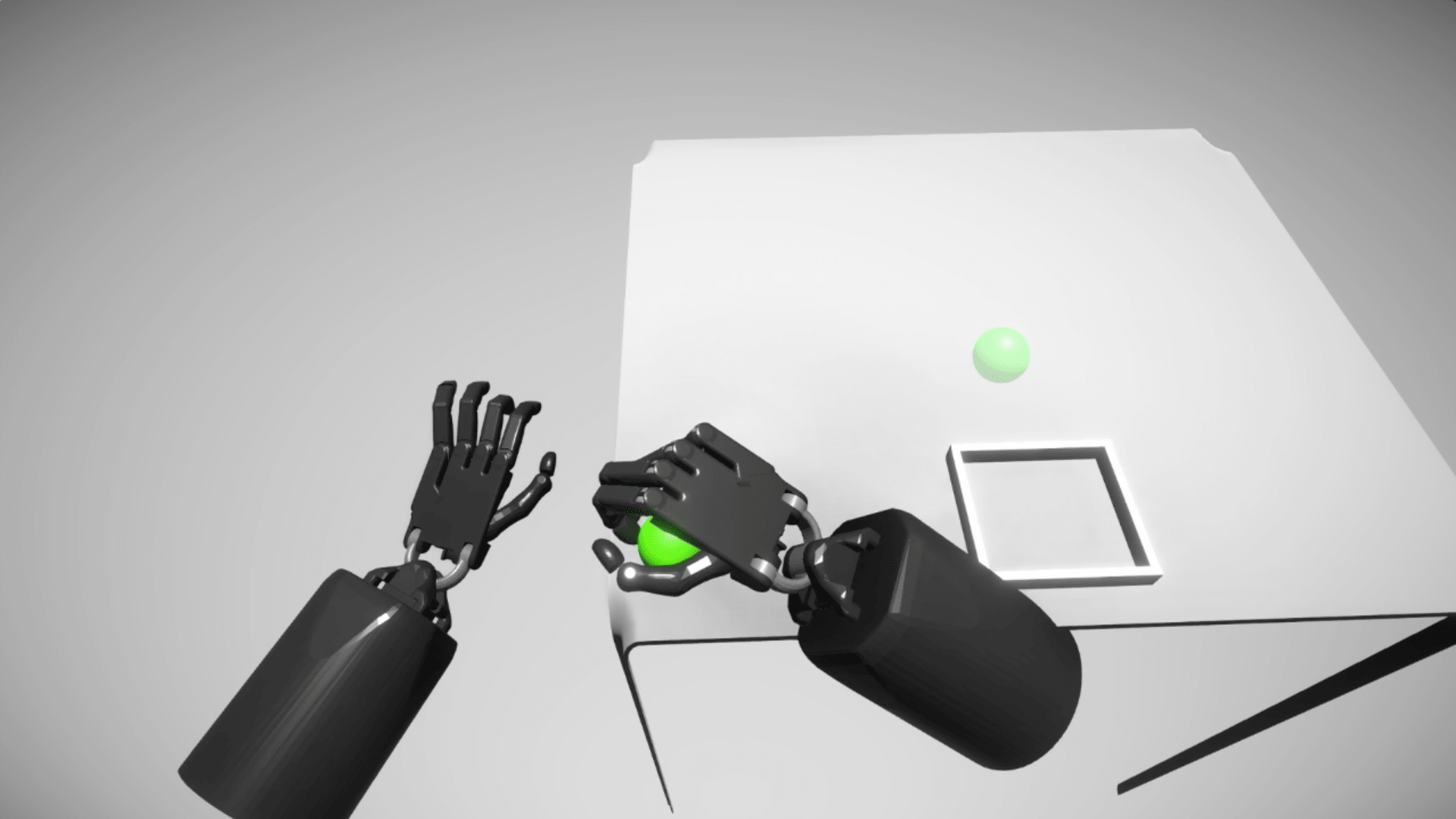}}
\subfigure{\label{fig:a7}\includegraphics[width=0.49\columnwidth]{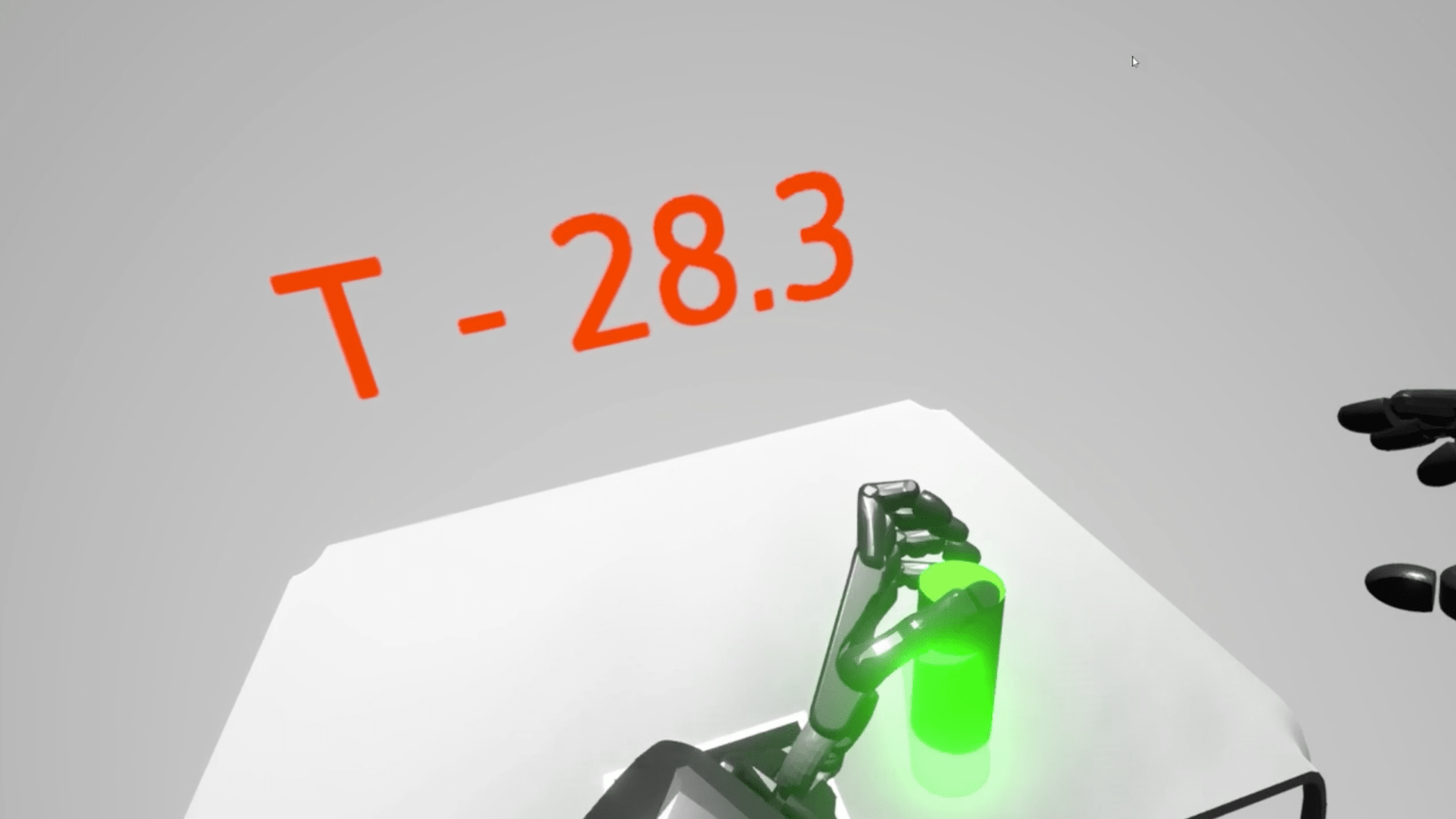}}
\subfigure{\label{fig:a8}\includegraphics[width=0.49\columnwidth]{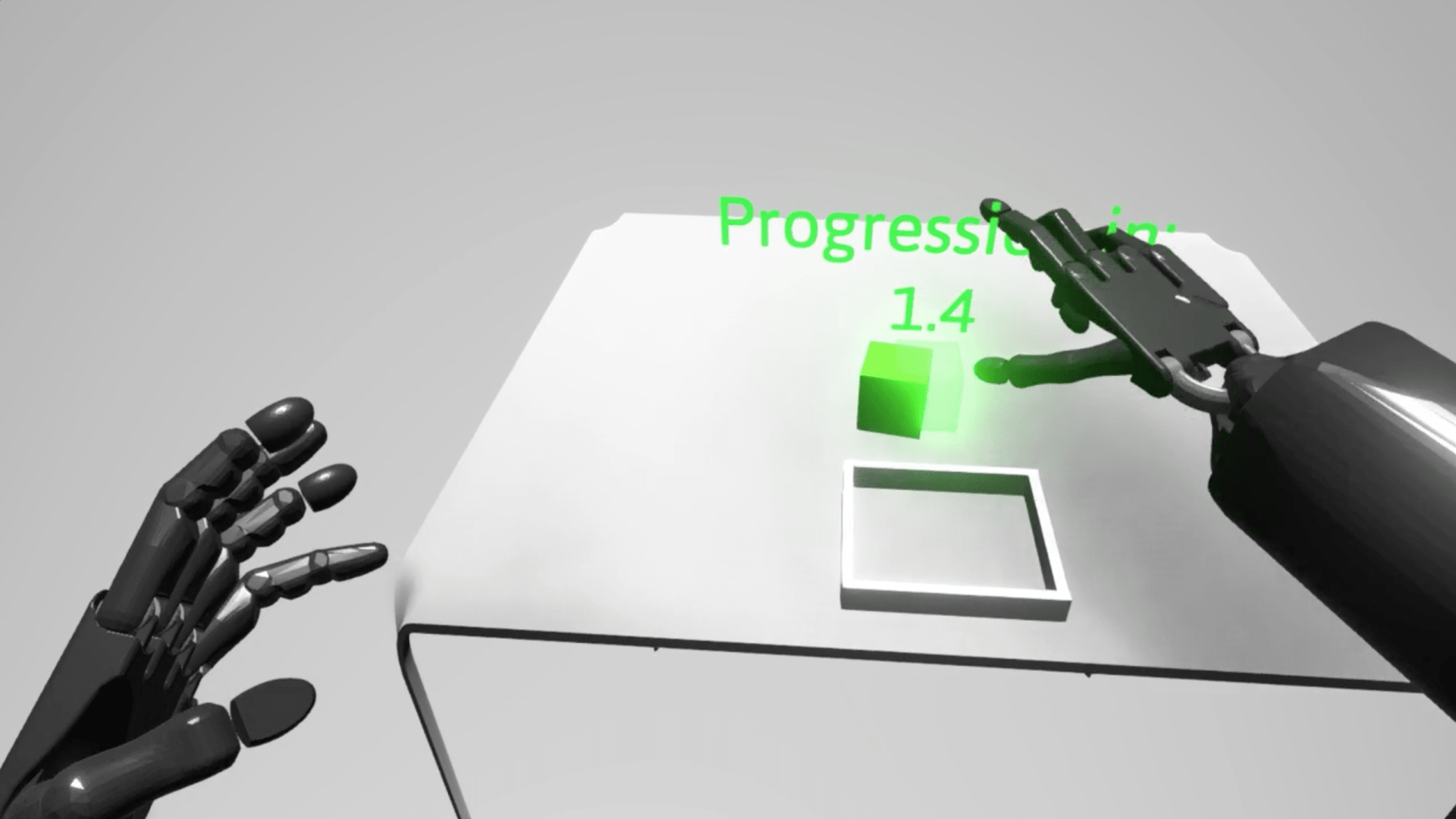}}
\vspace{-7.5mm}
\caption{Different participants during the manipulation experiment.}~\label{fig:ParticipantsExperiment}
\end{figure}

\section{Discussion}
Our results are summarised as follows: the overall performance of users increased by around 40\% by using stereoscopic vision with the VRHMD instead of monocular vision with the display monitor. 
Somatosensory feedback increased performance furthermore by 10\% over all measurements as well.
Auditory stimulation, however, had no significant effect on any measure apart from spatial accuracy, which increased by less than 5\%.

These results provide evidence to the untested hypothesis of \cite{Sigrist2013}. More specifically, our results show that an audiovisuohaptic interface, incorporating a stereoscopic VRHMD than a monocular monitor, contributes to the highest task performance, followed closely by visuohaptic and less closely by audiovisual interfaces. For a cone-like illustration of each interface effectiveness, that closely resemble the figures of \cite{Sigrist2013}, see the Appendices.

Our results support existing research that vision is the dominant sense \cite{Rock594,10.2307/40063340}, outperforming all other senses \cite{Heilig1992ELCD}.
As depth information is important in manipulation tasks, we can infer that better performance in VR may in part be due to the superior information available when using VRHMDs. 
This supports current literature
\cite{doi:10.1177/154193129503902006,Witmer:1998:JPT:1246749.1246755, 7164348}.

Our results showed that less perceived cognitive workload was observed in the use of the VRHMD than in the monocular display. 
This contradicts previous work \cite{doi:10.1177/1541931213601863}, but this may be attributed to significantly higher amounts of induced vection.
Thus full conclusions cannot be drawn with our static scenario and further investigation is required to confirm.
Our findings show that haptic feedback leads to better performance which is supported by some studies \cite{8446227}, but contradicts others \cite{Frid2018}.
The latter study found no significant effect of haptic feedback in a virtual throwing task. 
Since there are such a large number of options available for providing haptic feedback, findings may differ wildly simply by using a slightly different device.
More research may be needed to investigate how small variations in the way haptic feedback is delivered, affects performance and a standardised device may be needed to compare the actual effect of haptics on humans. 

The differences in the results for haptic devices may be partially explained by the "uncanny valley of haptics" \cite{BergerUncannyValleyHaptics}.
This suggests that increasing the resolution of haptic feedback without the corresponding level of stimulation from other senses, will not contribute to a guaranteed increase in performance.
Thus the resolution of all feedback interfaces has to be similar.
Their study \cite{BergerUncannyValleyHaptics} used handheld controllers to deliver haptic feedback. 
We used a custom vibrotactile glove which has a higher resolution than the handheld controllers, but this only increased performance when the resolution of visual stimulation was increased as well by switching from the monocular display monitor to the stereoscopic VRHMD, thus supporting \cite{BergerUncannyValleyHaptics}.

We found little evidence to show that auditory feedback has a positive impact on performance, though
spatial accuracy did increase in the audiovisual condition compared to the visual condition ($p<0.05$). Workload demand marginally decreased when auditory feedback was presented than just none at all, but not a significant level  ($p=0.056$). 
However, a significant difference was found in previous work \cite{nagai_kimura_tsuchiya_iida_2000}. 
It is possible that this was a bi-product of the increase in performance when switching from mono to stereo vision, potentially overshadowing the contribution of audio in the subjective performance of participants.

In both objective and subjective measures, the combination of stereoscopic visual feedback i.e. the VRHMD with the addition of audio and haptic feedback, Condition 8, provided the best performance overall. This supports our primary hypothesis. 
This is in line with existing literature, that adding more modalities is correlated to improved performance in manipulation scenarios \cite{Burke:2006:CEV:1180995.1181017}. Though, there was no significant difference in performance when using only two modalities: stereoscopic visual i.e. VRHMD and haptic feedback, Condition 6.
However, we did see a marginal, but still significant drop in position and orientation accuracy in this condition, indicating that auditory did contribute to the effectiveness of spatial accuracy.

The main findings and design implications of our study include: 
\begin{itemize}
\setlength
  \item Adding additional modalities increases performance
  \item Relying on just one modality should be avoided
  \item Vision dominates, making the highest contribution in performance when enhancing from mono to stereo vision
  \item Effectiveness of multi-modal interfaces is scenario-specific, this research explored it in the context of manipulation
  \item Prioritization of visual, somatosensory and then auditory stimulation should be given for manipulation scenarios
  \item Increasing task complexity lowers effectiveness as expected, but is not proportional for all multi-modal interfaces 
  \item Vibrotactile feedback can be considered as a low-cost somatosensory approach while more focus can be given on the design of vibrotactile intensity to compensate for the inherent lack of force-feedback
\end{itemize}

All of our hypotheses are summarized in \autoref{table:HypothesesSummary} below, providing an overall overview of our findings.

\begin{table}[H] \centering
\begin{footnotesize}
\begin{tabular}{lrrr} \toprule
 \textbf{Hypothesis} & & Support & Description \\ \midrule

\rowcolor{Gray}\begin{tabular}[c]{@{}l@{}}\textbf{H1:} Lower perceived workload\end{tabular}   & &
Partial
&
(a) Y (b) P (c) N\\ 

\begin{tabular}[c]{@{}l@{}}\textbf{H2:} Higher system usability\end{tabular}   & &
Partial
&
(a) Y (b) P (c) N\\  

\rowcolor{Gray}\begin{tabular}[c]{@{}l@{}}\textbf{H3:} Less task time\end{tabular}   & &
Partial
&
(a) Y (b) P (c) N\\  

\begin{tabular}[c]{@{}l@{}}\textbf{H4:} Less distance error\end{tabular}   & &
Partial
&
(a) Y (b) P (c) N\\ 

\rowcolor{Gray}\begin{tabular}[c]{@{}l@{}}\textbf{H5:} Higher placement precision\end{tabular}   & &
Full
&
(a) Y (b) Y (c) Y \\
\midrule
\multicolumn{4}{c}{(a) Vision with stereoscopic VR-HMD than monoscopic monitor} \\
\multicolumn{4}{c}{(b) Haptic feedback than without (c) Sound feedback than without} \\
\multicolumn{4}{c}{*P: Partial; only effective when paired with stereo VR-HMD} \\
\bottomrule
\end{tabular}
\end{footnotesize}
  \caption{Summary of Hypotheses support. Y: Yes, P: Partial, N: No.}
  \label{table:HypothesesSummary}
\end{table}

\subsection{Design and Research Implications}
Our low-cost haptic gloves show that expensive solutions are not required to achieve significant performance increases, in line with \cite{BergerUncannyValleyHaptics}.
This may enable a wider range of research into haptic feedback and cost-effective multi-modal interfaces.

We also show that adding haptic feedback to monocular feedback has no significant effect on performance.
However, adding haptic feedback to a VRHMD does improve performance significantly.
This seems to be in line with the "uncanny valley of haptics" \cite{BergerUncannyValleyHaptics}, which supports that it is not enough to add extra sensory modalities, but the resolution of these modalities must be similar.
This is highlighted in Conditions 2 (visuohaptic) \& Condition 4 (audiovisuohaptic), where monocular vision is used. In this case, the additional sensory modalities did not contribute to an observable difference in performance apart from spatial accuracy, possibly due to a mismatch in resolution between monocular vision and other modalities.

Priorities should be given when designing multi-modal interfaces for object manipulation. Our results support that researchers should aim to enhance visual stimuli before adding somatosensory feedback and lastly auditory. 

Furthermore, based on our results, designers and researchers focusing on human performance in teleoperation, are encouraged to combine sensory interfaces as highlighted in this study. 
We observed that almost in all cases, bi-modal feedback i.e. visuohaptic and even more so audiovisuohaptic interfaces are significantly better performing than just relying on visual feedback.
This may be even more the case for sensory channels that are already overloaded \cite{Lathan:2002:EOS:638095.638100, doi:10.1080/10447319609526138}, thus potentially opening more opportunities for researchers to investigate the effectiveness of such interfaces when channels are overloaded. 

\subsection{Limitations and Future Work}
The investigation of this research was focused on the contribution of the effectiveness of each sensory modality and combinations of these.
However, we have not yet tested how auditory or somatosensory feedback would have compensated potentially overloaded visual information, which would have provided furthermore insight. Furthermore, we investigated the effectiveness of common visual feedback modalities i.e. the monitor display monitor and a VRHMD with their inherent capabilities. However, we did not explicitly and strictly investigated how monocular and stereoscopic visual feedback by themselves would influence performance. Future road-map would include using the VRHMD with either monocular or stereoscopic rendering. In addition, multi-modal design decisions are of paramount importance before implementing any kind of sensory feedback \cite{Sigrist2013}. In our case, auditory feedback was implemented as the means of task indication and succession, instead of a continuous sonification i.e. concurrent type. Examples of concurrent auditory feedback would include controlling auditory pitch continuously based on target proximity, specific to manipulation tasks. Thus, further evidence may be needed on how not only different types of sensory feedback may influence task succession, but also how the design decisions of each sensory channel affect task efficiency. We did assume zero to minimal latency during our experiments, knowing that time delays are correlated to simulator sickness. This is a real-world problem in teleoperation and further aggravated in wireless technologies.
Latency in our experiments was <15ms and thus its effect was not studied.
However, in real-world applications, latency can become a problem that causes simulator sickness and is also a challenge in teleoperation where communication bandwidth is limited \cite{10.1145/3029798.3038369}.
Within this study, by thoroughly comparing an audiovisuohaptic multi-modal interface, we have gained interesting insight on which modalities contribute to increased task performance, as long as time-delay is minimal.

\section{Conclusion}
This paper explored how combining multiple sensory interfaces affects performance in manipulation tasks of varying complexity.
Each combination of visual (monocular display monitor or a stereoscopic VRHMD), audio (with or without) and haptic (with or without) interface was tested.
Task difficulty ranged from low to high by changing the size and shape of objects as well as distance to the target placement.

The performance was measured objectively and subjectively under experimental conditions.
The results of these experiments showed a 40\% increase in overall performance when using stereoscopic VRHMD visual feedback compared to a monocular display monitor.
Somatosensory stimulation contributed a furthermore 10\% increase in performance, while auditory feedback only increased spatial accuracy by an additional 5\%.

Our evaluation found that by adding one more sensory modality in an interface is of a significant benefit than just relying on visual feedback.
We thus conclude that task performance in teleoperation can be positively influenced by carefully selecting an appropriate combination of sensory feedback for a given task.
As a result of this study, future researchers and designers should identify and prioritize certain modalities when designing multi-modal interfaces.

\section*{ACKNOWLEDGMENT}
We would like to thank Prof. Taku Komura and Prof. Robert Fisher for approving the use of their facilities to accommodate the study. This project was funded by the EPSRC Future AI and Robotics for Space (EP/R026092/1).

\section*{APPENDIX}
\begin{figure*}[hbt!]
\centering
\subfigure{\label{fig:a11}\includegraphics[width=0.245\textwidth]{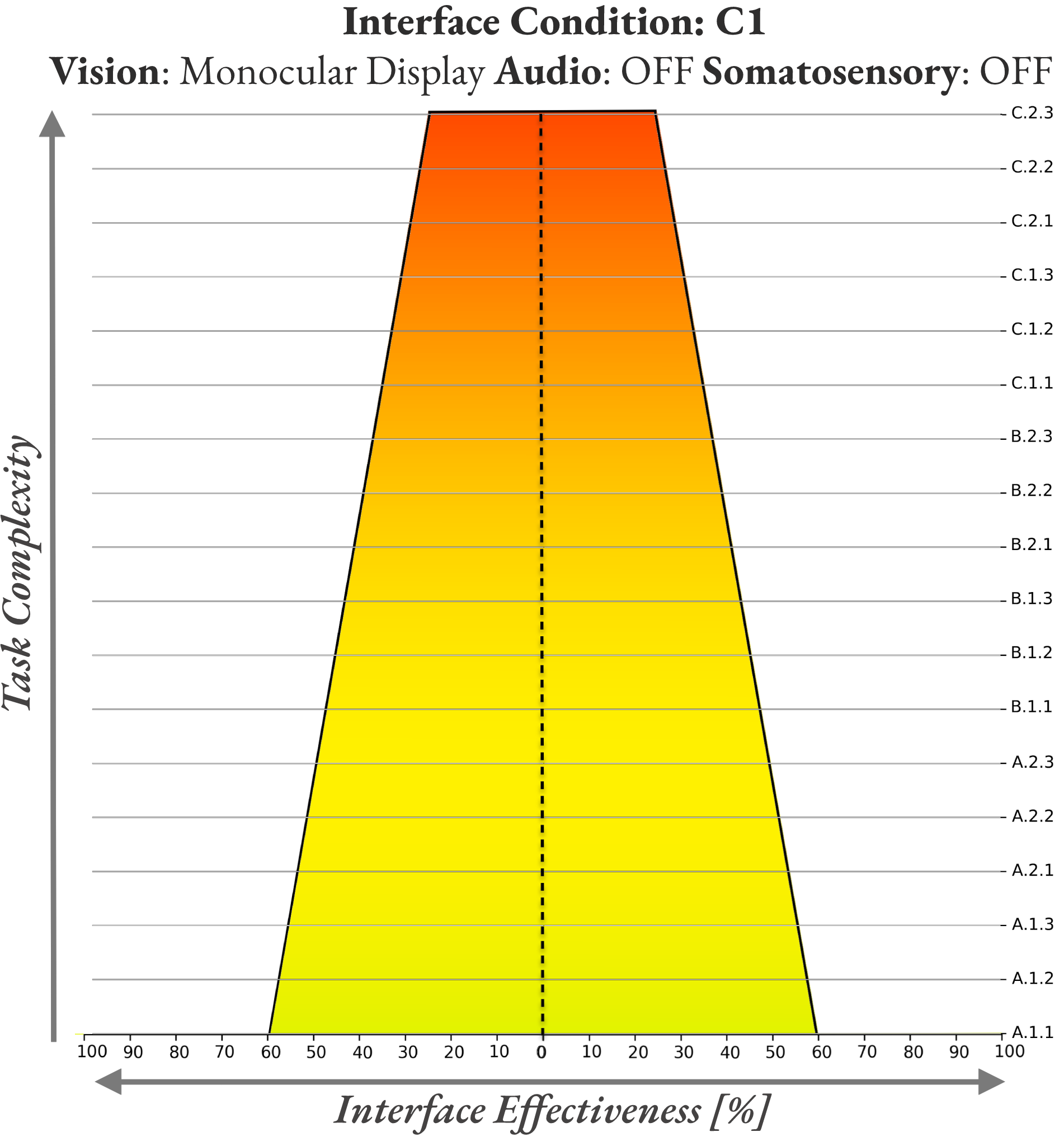}}
\subfigure{\label{fig:a22}\includegraphics[width=0.245\textwidth]{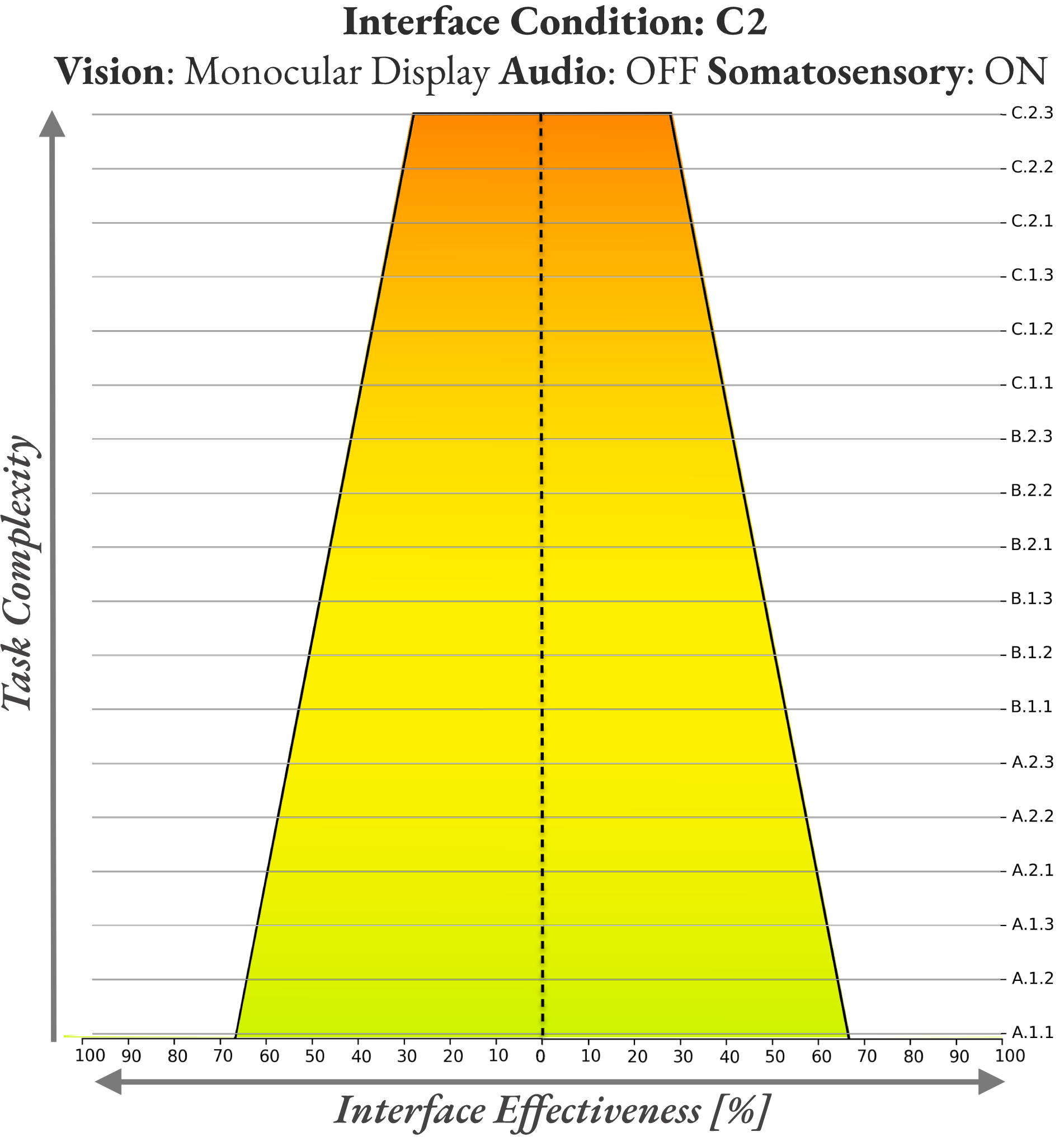}}
\subfigure{\label{fig:a33}\includegraphics[width=0.245\textwidth]{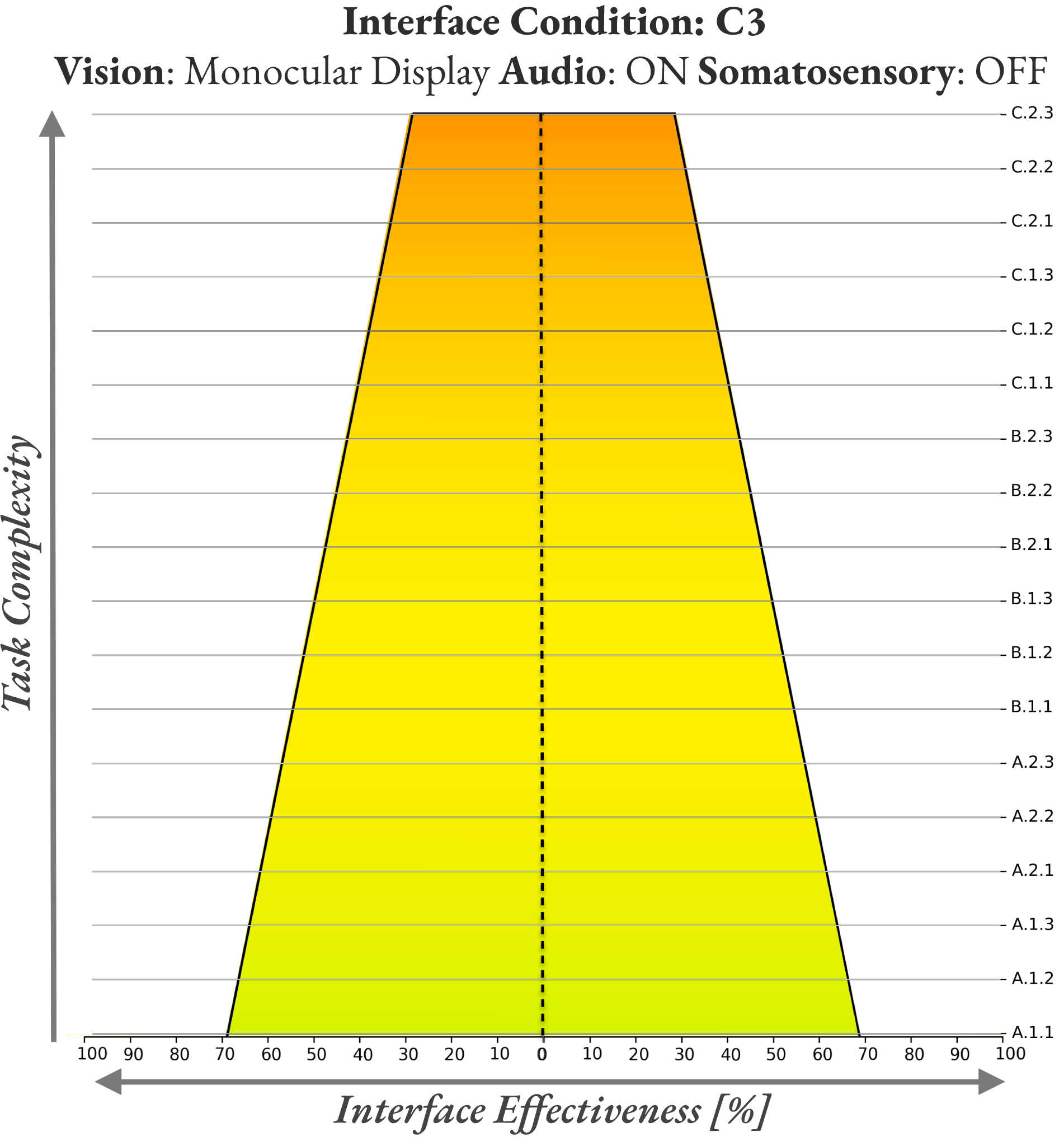}}
\subfigure{\label{fig:a44}\includegraphics[width=0.245\textwidth]{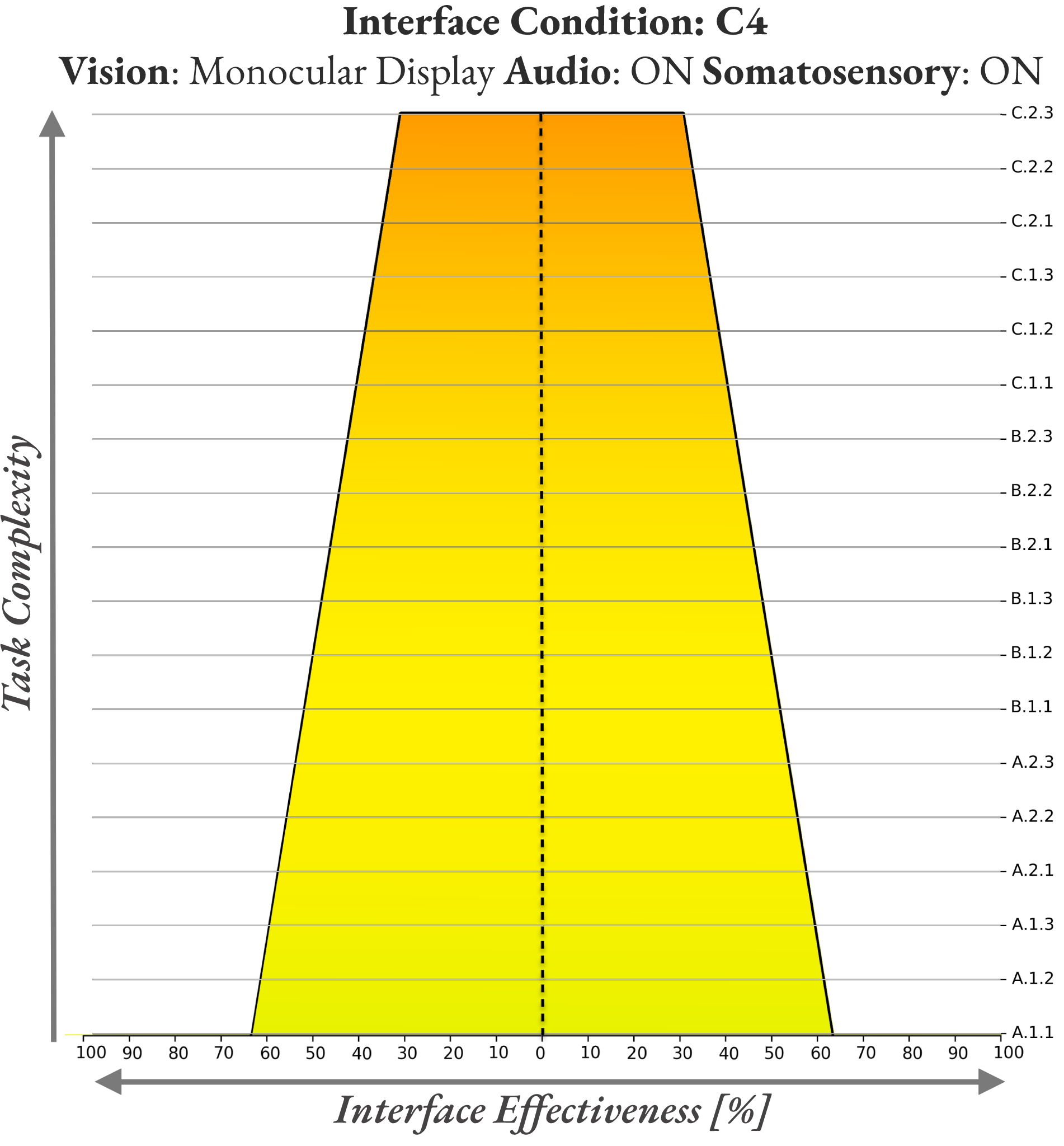}}
\subfigure{\label{fig:a55}\includegraphics[width=0.245\textwidth]{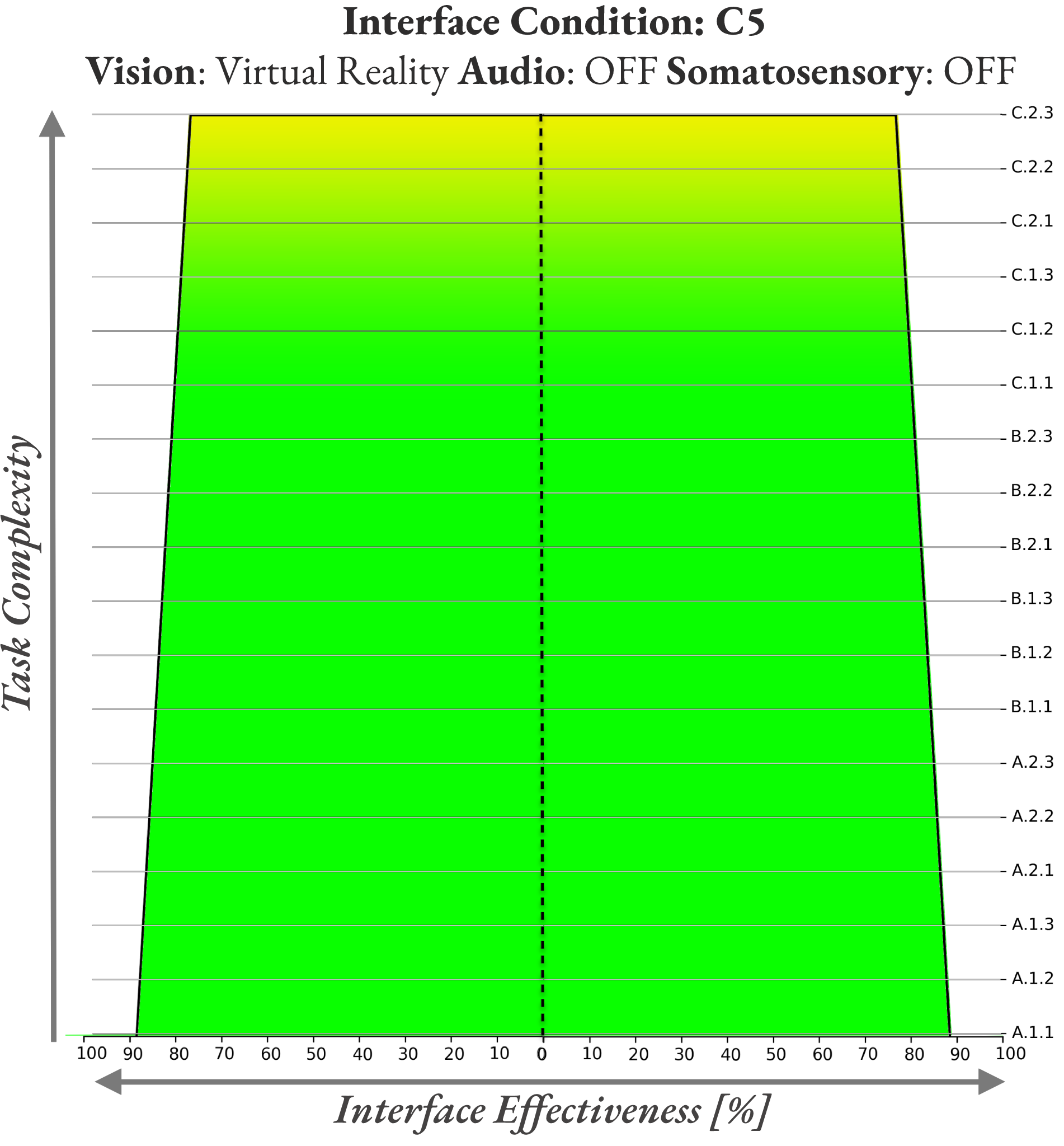}}
\subfigure{\label{fig:a66}\includegraphics[width=0.245\textwidth]{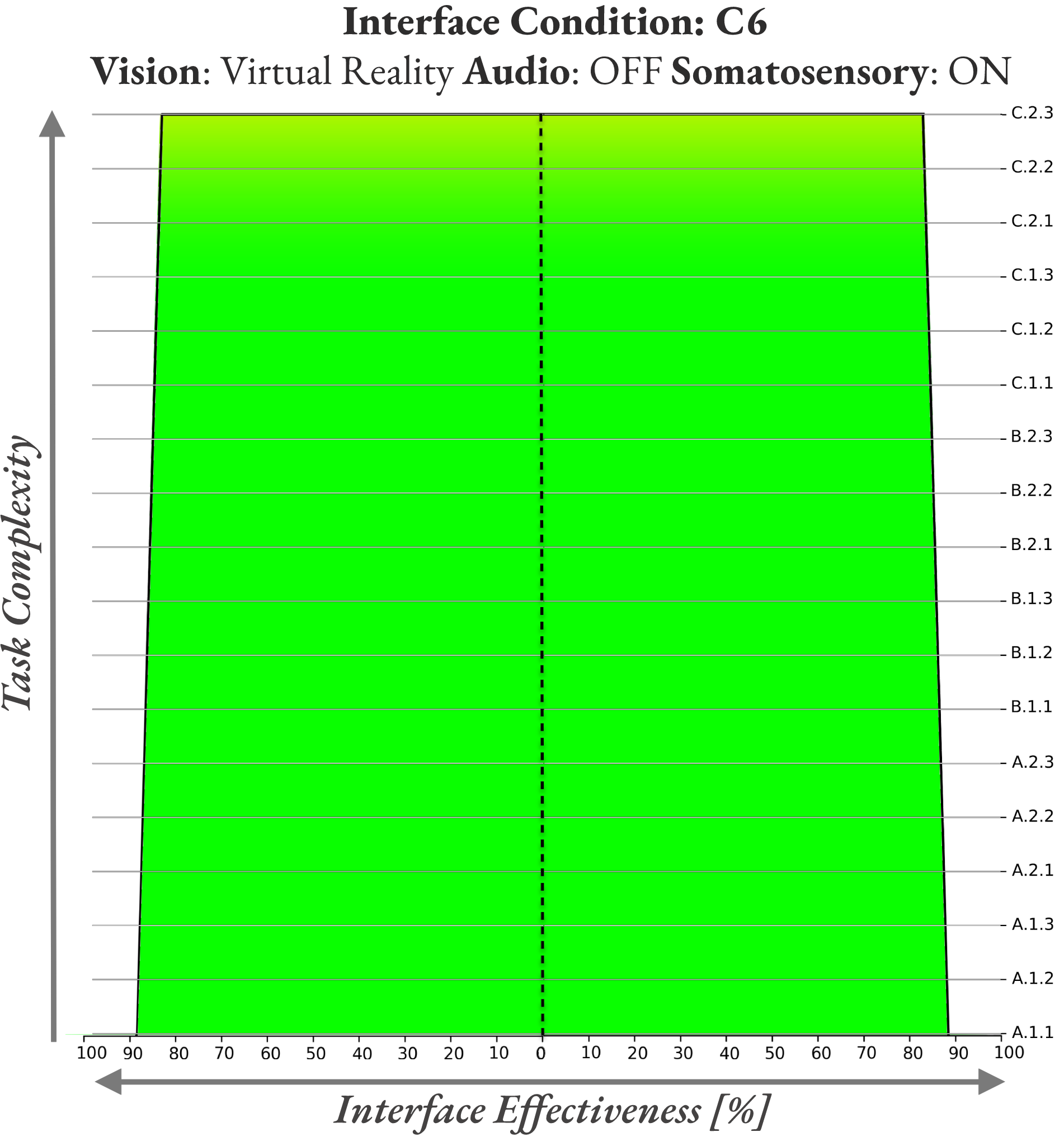}}
\subfigure{\label{fig:a77}\includegraphics[width=0.245\textwidth]{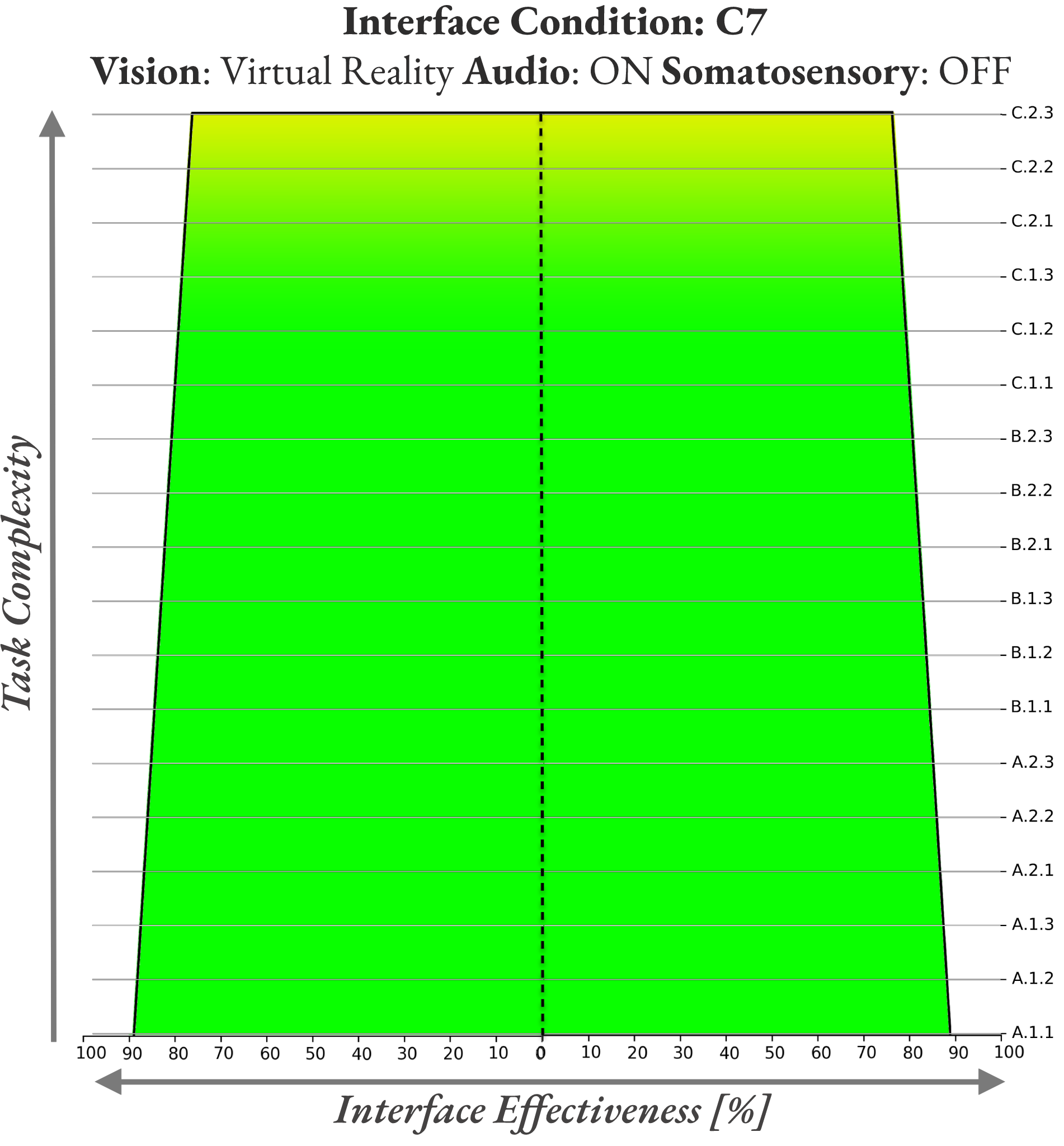}}
\subfigure{\label{fig:a88}\includegraphics[width=0.245\textwidth]{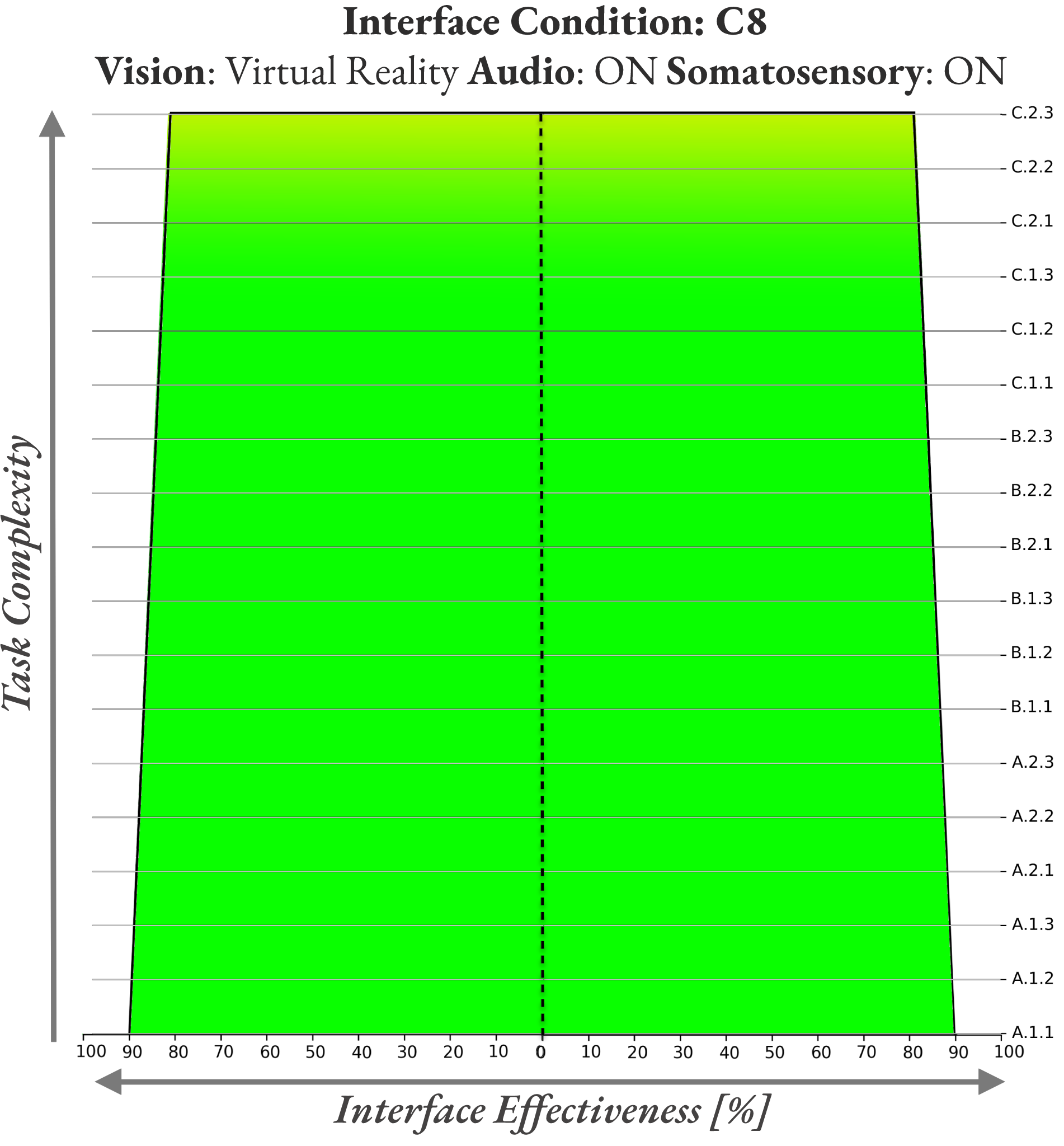}}
\caption{Overall interface effectiveness through linear regression, across all measurements and across all tasks with an increasing task complexity from lower to higher. Width of the shapes represents the effectiveness, the wider the higher. Colouring also indicates the effectiveness increasing from red to green. The overall effectiveness is calculated linearly, specifically, the measurements are weighted $\left ( 1 - {1}/{V_{max}}\right )$ where $V_{max}$ is the maximum limit of the measurement. The data points from the scatter plot have been line fitted through linear regression to visualize a cone-like illustration. The width of the cone represents the effectiveness while the height of the cone the effectiveness of the interface at the specific task complexity. The specific task complexity is discussed in section "Manipulation Tasks of Varying Complexity".}~\label{fig:TotalSummary}
\vspace{-5mm}
\end{figure*}

In this appendix section, we summarize the overall interface effectiveness from our experiments. We visualise the overall findings of our results in \autoref{fig:TotalSummary}. In these figures, we visualise the overall effectiveness of each individual interface condition across all measurements and all tasks thus giving the final overview of our entire experimental results. 

\bibliographystyle{IEEEtran}
\bibliography{main}
\end{document}